\documentclass[fleqn,usenatbib]{mnras}

\usepackage{newtxtext,newtxmath}
\usepackage[T1]{fontenc}
\usepackage{ae,aecompl}

\usepackage{graphicx}	% Including figure files
\usepackage{amsmath}	% Advanced maths commands
\usepackage{amssymb}	% Extra maths symbols
\usepackage{lipsum}
\usepackage{pdflscape}
\usepackage{afterpage}

\newcommand{\chandra}{\textit{Chandra}}
\newcommand{\nustar}{\textit{NuSTAR}}

\newcommand{\swift}{{\it Swift}}

\newcommand{\xmm}{{\it XMM-Newton}}

\newcommand{\hubble}{{\it Hubble Space Telescope}}

\newcommand{\Lya}{Ly$\alpha$}
\newcommand{\src}{Mrk~335}
\newcommand{\kms}{km~s$^{-1}$}

\title[The nuclear environment of Mrk~335]{The nuclear environment of the NLS1 Mrk 335: obscuration of the X-ray line emission by a variable outflow}

\author[M. L. Parker et al.]{M. L. Parker,$^{1}$\thanks{E-mail: mparker@sciops.esa.int}
A. L. Longinotti,$^{2,3}$
N. Schartel,$^{1}$
D. Grupe,$^4$
S. Komossa,$^5$
G. Kriss,$^{6}$\newauthor
A. C. Fabian,$^7$
L. Gallo,$^8$
F. A. Harrison,$^{^9}$
J. Jiang,$^{7,10}$
E. Kara,$^{11}$
Y. Krongold,$^{12}$\newauthor
G. A. Matzeu,$^{1}$
C. Pinto,$^{13}$
M. Santos-Lle\'{o},$^{1}$
\\
$^{1}$European Space Agency (ESA), European Space Astronomy Centre (ESAC), E-28691 Villanueva de la Ca\~{n}ada, Madrid, Spain\\
$^{2}$Instituto Nacional de Astrof\'{i}sica, \'{O}ptica y Electr\'{o}nica, Luis E. Erro 1, Tonantzintla, Puebla, C.P. 72840, Mexico\\
$^{3}$CONACyT-INAOE\\
$^{4}$Department of Physics, Earth Sciences, and Space System Engineering, Morehead State University, Morehead, KY 40514, USA\\
$^{5}$Max-Planck-Institut f\"{u}r Radioastronomie, Auf dem H\"{u}gel 69, 53111 Bonn, Germany\\
$^{6}$Space Telescope Science Institute, 3700 S. Martin Drive, Baltimore, MD 21218, USA\\
$^{7}$Institute of Astronomy, University of Cambridge, Madingley Road, Cambridge, CB3 0HA, UK\\
$^{8}$Department of Astronomy and Physics, Saint Mary's University, Halifax, Nova Scotia, Canada\\
$^{9}$Cahill Center for Astrophysics, 1216 E. California Blvd, California Institute of Technology, Pasadena, CA 91125, USA\\
$^{10}$Tsinghua Center for Astrophysics, Tsinghua University, Beijing 100084, China\\
$^{11}$Department of Astronomy, University of Maryland, College Park, MD 20742-2421, USA\\
$^{12}$Instituto de Astronom\'{i}a, Universidad Nacional Aut\'{o}noma de M\'{e}xico, Circuito Exterior, Ciudad Universitaria, Ciudad de M\'{e}xico 04510, M\'{e}xico\\
$^{13}$European Space Agency, European Space Research \& Technology Centre (ESTEC), Postbus 299, 2200 AG Noordwijk, The Netherlands\\
}

\date{Accepted XXX. Received YYY; in original form ZZZ}

\pubyear{2019}

\begin{document}
\label{firstpage}
\pagerange{\pageref{firstpage}--\pageref{lastpage}}
\maketitle

\begin{abstract}
We present \xmm , \nustar , \swift\ and \hubble\ observations of the Narrow-line Seyfert 1 galaxy Mrk~335 in a protracted low state in 2018 and 2019. The X-ray flux is at the lowest level so far observed, and the extremely low continuum flux reveals a host of soft X-ray emission lines from photoionised gas. The simultaneous UV flux drop suggests that the variability is intrinsic to the source, and we confirm this with broad-band X-ray spectroscopy. The dominance of the soft X-ray lines at low energies and distant reflection at high energies, is therefore due to the respective emission regions being located far enough from the X-ray source that they have not yet seen the flux drop.
Between the two \xmm\ spectra, taken 6 months apart, the emission line ratio in the O\textsc{vii} triplet changes drastically. We attribute this change to a drop in the ionisation of intervening warm absorption, which means that the absorber must cover a large fraction of the line emitting region, and extend much further from the black hole than previously assumed. The HST spectrum, taken in 2018, shows that new absorption features have appeared on the blue wings of  C\textsc{iii}$^{*}$, \Lya , N\textsc{v}, Si\textsc{iv} and C\textsc{iv}, likely due to absorbing gas cooling in response to the low flux state.
\end{abstract}

\begin{keywords}
galaxies: active -- accretion, accretion disks -- black hole physics -- X-rays: individual: Mrk 335
\end{keywords}

%%%%%%%%%%%%%%%%%%%%%%%%%%%%%%%%%%%%%%%%%%%%%%%%%%

%%%%%%%%%%%%%%%%% BODY OF PAPER %%%%%%%%%%%%%%%%%%

\section{Introduction}

% \red{Soft emission in AGN}
% \red{Type 2 stuff.} Bianchi 06. Kinkhabwala.
Since the launch of \xmm\ and \chandra , soft X-ray emission lines from photoionised gas have been observed in many AGN. These observations primarily take place in type 2 AGN, where the heavy obscuration of the primary continuum means that the direct AGN continuum is suppressed and emission from the nuclear environment can be seen \citep[e.g.][]{Sako00_mrk3, Sambruna01, Kinkhabwala02, Ogle03, Bianchi06}. It is generally thought that type 1 and 2 AGN are the same sources, but viewed from different angles \citep[the unified model:][]{Urry95}, where the obscuration seen in type 2 AGN is simply out of the line of sight in type 1s. The same photoionised emission lines should therefore be present in Seyfert 1 galaxies. This is more difficult to observe, due to the much larger contribution of the continuum emission, and the AGN nuclear environment is more commonly glimpsed through absorption. Nevertheless, with high signal spectra, many emission lines have been observed in type 1 AGN \citep[e.g.][]{Kaspi02,Reeves16,Behar17}

One way to avoid this problem of contamination from the continuum is to observe type 1 AGN at low flux levels, and this approach has been productive in the past for sources where the continuum flux has dropped sharply \citep[e.g.][]{Longinotti08, Peretz19} and where the continuum has been obscured, allowing the soft lines to be observed \citep[e.g.][]{Armentrout07,Nucita10,Whewell15, Mao19_ngc3783}. However, even the best observations of low-flux type 1 AGN to date generally only yield a static snapshot of the scattering gas, where it is hard to separate different emission and absorption regions.

Nevertheless, there have been a handful of studies aimed at measuring variability of soft emission lines in AGN. \citet{Detmers09} tracked the flux of the O\textsc{vii} forbidden line between 7 high resolution observations of NGC~5548, using the variability timescale of the line to constrain the geometry of the BLR. \citet{Landt15a, Landt15b} combined observations of the O\textsc{vii} in X-rays with [Fe\textsc{vii}] in the optical to estimate the density and hence location of the coronal emission lines in NGC~4151 and NGC~5548.

% \red{Mrk 335}
Mrk~335 is one of the best known and studied NLS1 galaxies, famous for changing in flux by orders of magnitude, particularly in the X-ray band \citep{Grupe07}.
The low flux intervals of Mrk~335 have been studied in detail in the past \citep{Grupe08, Grupe12}, and show very rich spectra with strong soft emission lines \citep{ Longinotti08}, complex absorption \citep{Longinotti13, Longinotti19}, and evidence for strong relativistic reflection from the inner accretion disk \citep{Gallo13, Gallo15, Gallo19, Parker14_mrk335, Wilkins15_mrk335}. While NLS1s generally have high star formation rates, Mrk~335 shows a very low contribution from star formation \citep{Sani10}. This means that Mrk~335 is the ideal source for studying the nuclear environment without contamination from a nuclear starburst.

In this paper, we report on the lowest flux observations so far taken of Mrk~335, where the continuum flux has dropped dramatically, revealing a host of emission lines at low energies. 

%%%%%%%%%%%%%%%%%%%%%%%%%%%%%%%%%%%%%%%%%%%%%%%%%%

\section{Observations and data reduction}

In July 2018 \emph{Neil Gehrels Swift Observatory} (hereafter \swift ) monitoring revealed that the UV flux from \src\ was the lowest ever observed, while the X-ray flux was also extremely low \citep{Grupe18_335Atel1,Grupe18_335Atel2}. Based on this, we triggered an approved \xmm/\nustar\ ToO (PI Schartel) for observations of AGN in low flux states. After 6 months, \src\ remained in the same low state, so we requested and were awarded an additional ToO observation (PI Parker) with \xmm\ to search for variability of the soft emission lines.

The details of the \xmm , \nustar\ and HST observations are given in Table~\ref{ObsTbl}.

\subsection{XMM-Newton}

Mrk~335 was observed twice with \xmm\ \citep{Jansen01}, the first time on 2018-07-11 for 114.5~ks (obs-ID 0780500301) and the second on 2019-01-08 for 117.8~ks (obs-ID 0831790601), with net 0.5--10~keV EPIC-pn count rates of 0.47~s$^{-1}$ and 0.40~s$^{-1}$.

We reduce the \xmm\ data using the Science Analysis Software (SAS) version 18.0.0, with the latest calibration files as of June 2019. We reduce the EPIC data with the \textsc{epproc} and \textsc{emproc} tools, and extract source photons from a 20$^{\prime\prime}$ radius circular region centered on the source, and background photons from a large circular region nearby, avoiding areas of high instrumental copper background and other sources. After filtering for background flaring with a threshold of 0.5~counts~s$^{-1}$ in the 10--12~keV band, the clean exposures are 91~ks for obs-ID 0780500301 and 68~ks for 0831790601, which is heavily affected by flaring towards the end of the observation.
We bin the EPIC spectra to oversample the instrumental resolution by a factor of 3 and to a minimum signal to noise ratio of 6.
For simplicity, we consider only the high signal-to-noise EPIC-pn data unless otherwise stated.

We reduce the Reflection Grating Spectrometer (RGS) data using the \textsc{rgsproc} tool, filtering for background flares. We combine the first order spectra from both instruments into a single spectrum, and bin to a constant 8 channels per bin.

\subsection{NuSTAR}
One \nustar\ \citep{Harrison13} exposure of 82~ks was taken 2018-07-10, overlapping with the \xmm\ observation. We reduce the data and extract spectra for the two Focal Plane Modules (FPMA and FPMB) using the \nustar\ Data Analysis Software (NuSTARDAS), included in NASA's High Energy Astrophysics Software (HEASOFT version 6.25), with CALDB version 20190513. We  extract source photons from a 30$^{\prime\prime}$ radius circular region centred on the source, and background photons from a large circular region on the same chip, avoiding the chip gap and other sources. 
We bin the FPM spectra to oversample the instrumental resolution by a factor of 3 and to a minimum signal to noise ratio of 6. For plotting purposes, we group the spectra in \textsc{xspec} for clarity, but fit them separately.

\subsection{Swift}
Since the discovery of Mrk 335 in its extreme low state in 2007 by \swift\ \citep{Grupe07} we have monitored this NLS1 on a regular basis (PI Grupe). Most of the \swift\ X-ray Telescope observations \citep[XRT,][]{Burrows05} were performed in photon counting mode \citep[pc-mode][]{hill04}. After processing the data with the Ftool {\it xrtpipeline} we extracted the spectra and event files with the tool {\it xselect}.
The circular source extraction region was centered on the position of Mrk 335 with and extraction radius of 58$^{''}$. Background counts were extracted from a nearby circular region with a radius of 235$^{''}$.  %Hardness ratios\footnote{here we define the hardness ratio as HR = $\frac{H-S}{H+S}$ where S and H are the counts in the 0.3-1.0 keV and 1.0-10.0 keV band, respectively.} were determined by the Bayesian method described by \citet{park06}.

The data taken by the UV/Optical telescope on board \swift\ \citep[UVOT, ][]{roming05} were co-added in each filter by the UVOT tool {\it uvotimsum}. The source counts were selected in a circle with a radius of 5$^{''}$ and 20$^{''}$ for the background region. Fluxes and magnutudes were determined using the UVOT task {\it uvotsource} applying the latest calibration files as described by \citet{poole08} and \citet{breeveld10}. The UVOT data were corrected for Galactic reddening \citep[$E_{B-V}=0.035$, ][]{sfd98} using the standard Galactic reddening curved given by \citet{cardelli89}. The equations to calculate the correction factors in each UVOT filter are given in \citet{roming09}.

\subsection{HST}
Upon detection of the low-flux state in Mrk 335, we also triggered
observations using the Cosmic Origins Spectrograph (COS) \citep{Green12}
on the {\it Hubble Space Telescope} (HST; PI Grupe).
We used gratings G130M and G160M to cover the wavelength range from
1132--1802 \AA\ with a resolving power of $\sim15,000$.
Each exposure used four focal-plane positions to enable removal of
detector artifacts.
The individual exposures were calibrated and combined using v3.3.4 of the COS
calibration pipeline, CALCOS.

\begin{table*}
  \centering
        \caption{Details of the observations of Mrk 335}
        \label{ObsTbl}
\begin{tabular}{l l c c c c}
\hline\hline
Observatory &Data Set Name/Obs. ID & Grating/Tilt  & Date & Start Time & Exposure Time\\
       &       &               &      &    (GMT)   & (s)\\
\hline
HST COS & ldrv01010 & G130M/1291 & 2018-07-23 &  22:22:50 &  2100 \\
& ldrv01020 & G160M/1623 & 2018-07-24 &  00:16:16 &  2320 \\
\xmm\   & 0780500301 & & 2018-07-11 & 01:26:22  & 114500 \\
        & 0831790601 & & 2019-01-08 & 12:04:00  & 117800 \\
\nustar & 80201001002 & & 2018-07-10 & 17:31:09 & 82257 \\
\hline
\end{tabular}

\end{table*}

%%%%%%%%%%%%%%%%%%%%%%%%%%%%%%%%%%%%%%%%%%%%%%%%%%%%%%%%%%%%%%%%%%%%%%%%%%%%%%%%%%%%%%%%%%%%%%%%%%%%

\section{Results}

\subsection{Long-term Lightcurve}

\begin{figure*}
    \centering
    \includegraphics[width=\linewidth]{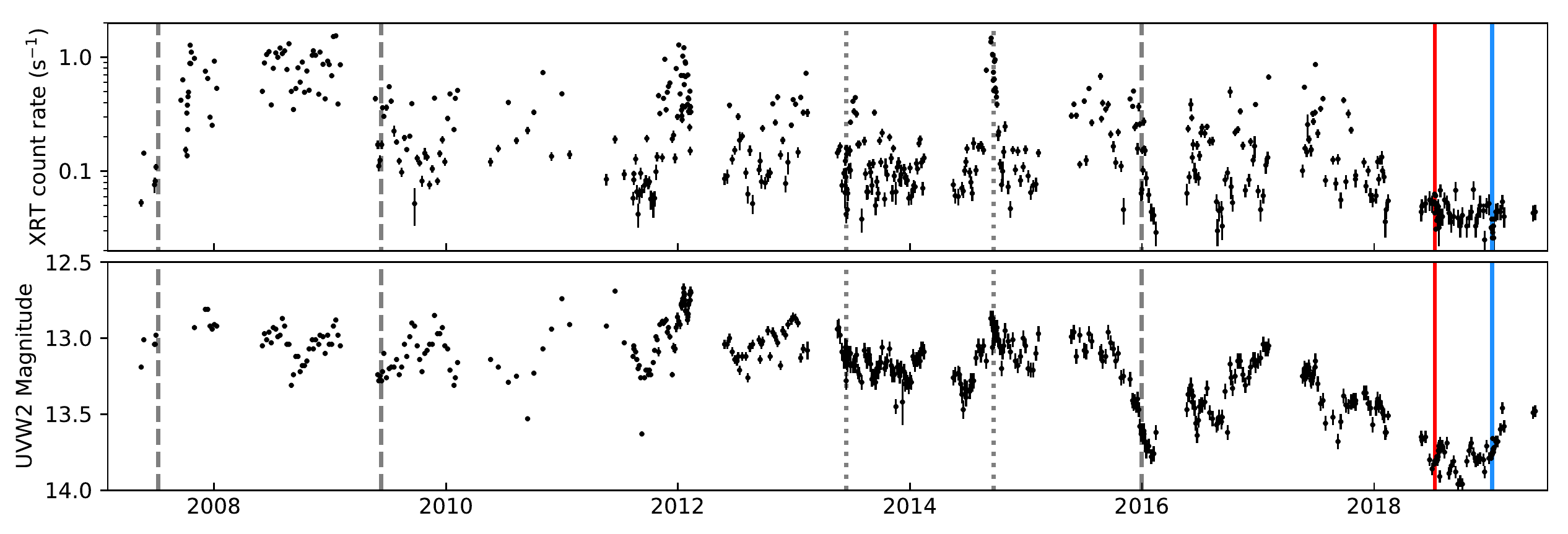}
    \caption{Long-term \swift\ lightcurve of \src , in X-rays and UV. At the time of writing, the source has been in an extended low flux interval, lasting approximately a year. Vertical lines indicate the dates of previous observations with \xmm\ (dashed) and \nustar\ (dotted). The 2018 \xmm/\nustar\ and 2019 \xmm\ observations are marked with red and blue solid lines, respectively.}
    \label{fig:lightcurve}
\end{figure*}

The long-term X-ray and UVW2 lightcurve (central wavelength of 1928~\AA) from \swift\ is shown in Fig.~\ref{fig:lightcurve}, with the dates of major observations marked. For the past year, both the X-ray and UV fluxes have been at historically low levels, and a protracted low flux state like this has never before been observed in this source. 

It is noteworthy that the X-ray flux is less variable than the UV in this low state. As we discuss later in the paper, this is most likely due to the X-ray spectrum being dominated by reprocessed emission, so the continuum variability can no longer be detected.

\subsection{HST}
\label{sec_hst}

Our HST spectra cover the most prominent UV resonance lines in Mrk 335, Ly $\alpha$, \ion{N}{v} $\lambda\lambda1238,1242$, \ion{Si}{iv} $\lambda\lambda1393,1403$, and \ion{C}{iv} $\lambda\lambda1548,1550$. Strong, broad, blueshifted absorption is visible associated with each of these resonance lines as well as in the lower-ionisation excited-state transition
\ion{C}{iii}* $\lambda1176$. To measure the strength of these absorption troughs, we fit an emission model to the lines and continuum of Mrk 335 that is identical in form to that used by \citet{Longinotti19}. We allowed the strengths of each emission component to vary freely to match the overall reduction in flux in Mrk 335, and we excluded the regions affected by broad absorption on the blue wings of all the emission lines.
% Figure \ref{fig:HST_CIV} shows the data and the resulting fit to the spectrum in the region surrounding the \ion{C}{4} emission line.

\begin{figure*}
    \centering
    \includegraphics[width=14cm]{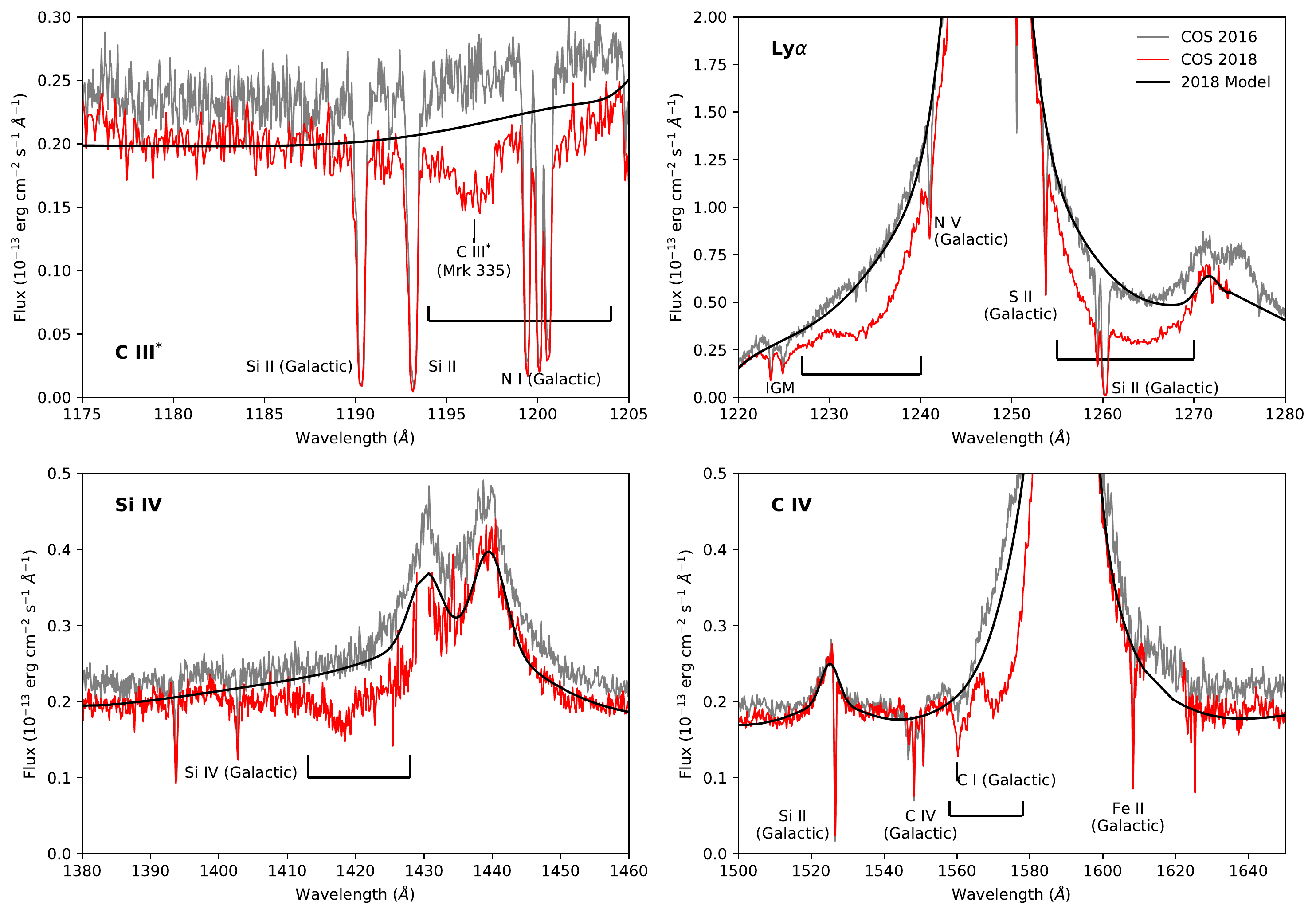}
    \caption{HST spectra of Mrk~335 in 2016 and 2018, showing the appearance of broad blueshifted absorption troughs on the C\textsc{iii}$^{*}$, Ly$\alpha$, Si\textsc{iv} and C\textsc{iv} lines, and absorption from \ion{N}{v} on the red wing of the \Lya\ line. The solid black lines show the emission model for the 2018 data, with the absorption features not modelled. A slight Gaussian smoothing is applied to the data in each case for clarity. Horizontal bars indicate the position of the broad absorption features. }
    \label{fig:hst}
\end{figure*}

\begin{figure}
    \centering
    \includegraphics[width=\linewidth]{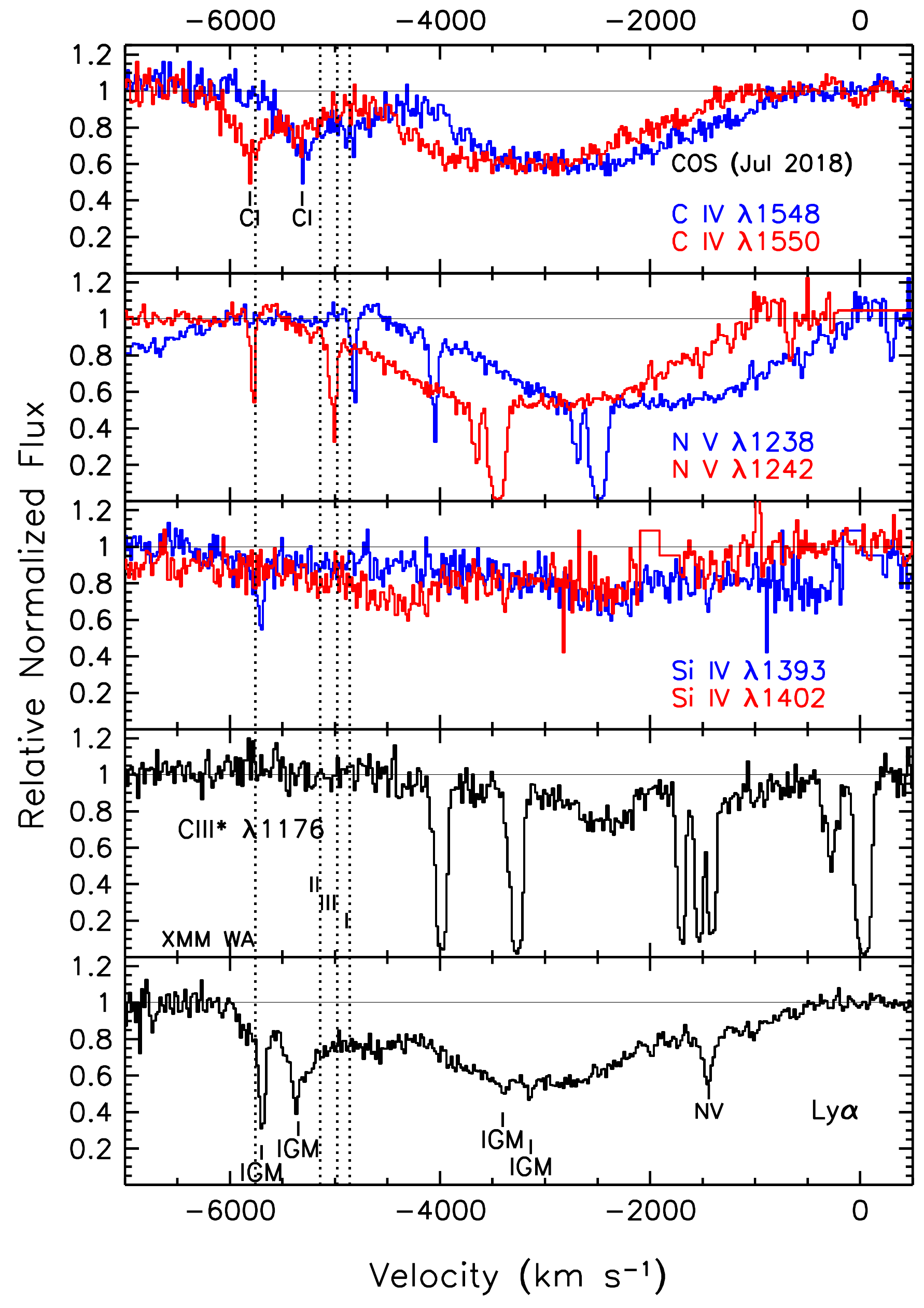}
    \caption{Normalised HST COS spectra of Mrk~335 in 2018 in velocity space, showing the new absorption features on the blue sides of the emission lines. Dashed vertical lines indicate the velocities of previously detected X-ray absorption \citep{Longinotti13,Longinotti19}}
    \label{fig:hstnorms}
\end{figure}

In Fig.~\ref{fig:hst} we show the line profiles and resulting fits. Clear broad absorption features are present on the blue sides of the lines, consistent with a drop in the ionisation of the absorbing gas, caused by the low source flux, making the observed UV absorption lines stronger. We discuss this further in the context of the X-ray spectra in section~\ref{sec_broadband}.
In Fig.~\ref{fig:hstnorms} we show the normalised COS spectra in velocity space. The broad absorption is clearly visible in each case, with a typical velocity of $\sim3000$~km~s$^{-1}$.

A full analysis of the HST data including detailed modelling of the absorption will be presented in a future paper (Grupe et al., in prep).

\subsection{RGS Spectrum}
\label{sec_rgs}

We initially fit just to the RGS spectrum, to establish some baseline properties of the warm absorption and photoionised emission before moving on to the broad-band fit.
By eye, there are striking differences in the emission lines between the two observations (Fig.~\ref{fig:rgs_spectra}). Most obviously, the O\textsc{vii} triplet in 2018 is dominated by the resonance line, while in 2019 the forbidden line dominates. Additionally, the O\textsc{vii} 1s-3p line observed at $\sim$19~\AA\ is weaker in 2019, and the C\textsc{vi} radiative recombination continuum (RRC) at $\sim$26~\AA\ is also weaker. The O triplet lines are usually parameterised by $R=z/(x+y)$ and $G=(z+x+y)/w$, where $w$ is the resonance line strength, $x+y$ the intercombination lines, and $z$ the forbidden line \citep{Porquet00}. Fitting these lines with Gaussians plus a power-law continuum in the two spectra gives ratios of $R=1.89$ and 3.06 for 2018 and 2019, respectively, and $G=1.7$ and 3.8. As shown in Fig.~8 of \citet{Porquet00}, if taken at face value this implies both a sharp drop in the temperature of the emitting gas and an order of magnitude drop in the density. Given the overall similarity of the two spectra, and the short interval between the observations, these kind of large scale changes in the nature of the photoionised gas are extremely unlikely, so this cannot be the cause of the change in line ratios. 

In addition to the dominant variability of the resonance line, there is possibly also some variability of the forbidden line. However, this is exaggerated in Fig.~\ref{fig:rgs_spectra}, as there is a single broad bin on the red side of the forbidden line which weakens the peak flux of the line. From the fit with Gaussians, the normalisations of the forbidden line are $4.2\pm0.7$~photons~cm$^{-2}$~s$^{-1}$ in 2018, and $4.7\pm0.5$~photons~cm$^{-2}$~s$^{-1}$ in 2019, so the change is not significant.
We also investigate the line ratios in the \ion{Ne}{ix} triplet, however the constraints are much weaker. In particular, the intercombination line is extremely uncertain in both cases, and only an upper limit on the line strength is found in the 2018 spectrum. As such, the ratios are consistent with those found from \ion{O}{vii}, but they are essentially unconstrained. 

To examine the changes in spectral shape further, we switch from fitting phenomenological Gaussians to full physical modelling.
To fit the emission lines, we use a \textsc{xspec} table version of the \textsc{spex} \citep{Kaastra96} photoionised emission/absorption model \textsc{pion} \citep{Miller15_pion}. This model self-consistently computes the ionisation equilibrium from the input continuum, and can be used to model both emission and absorption features, although in this case we use it exclusively for the emission. We calculate a grid of photoionised emission spectra using \textsc{pion}, and convert this into an \textsc{xspec} table, as discussed in Appendix~\ref{sec_pionxs}. To distinguish it from the original, analytical version of \textsc{pion}, we refer to this model as \textsc{pion\_xs}.

\begin{figure*}
    \centering
    \includegraphics[width=0.3\linewidth]{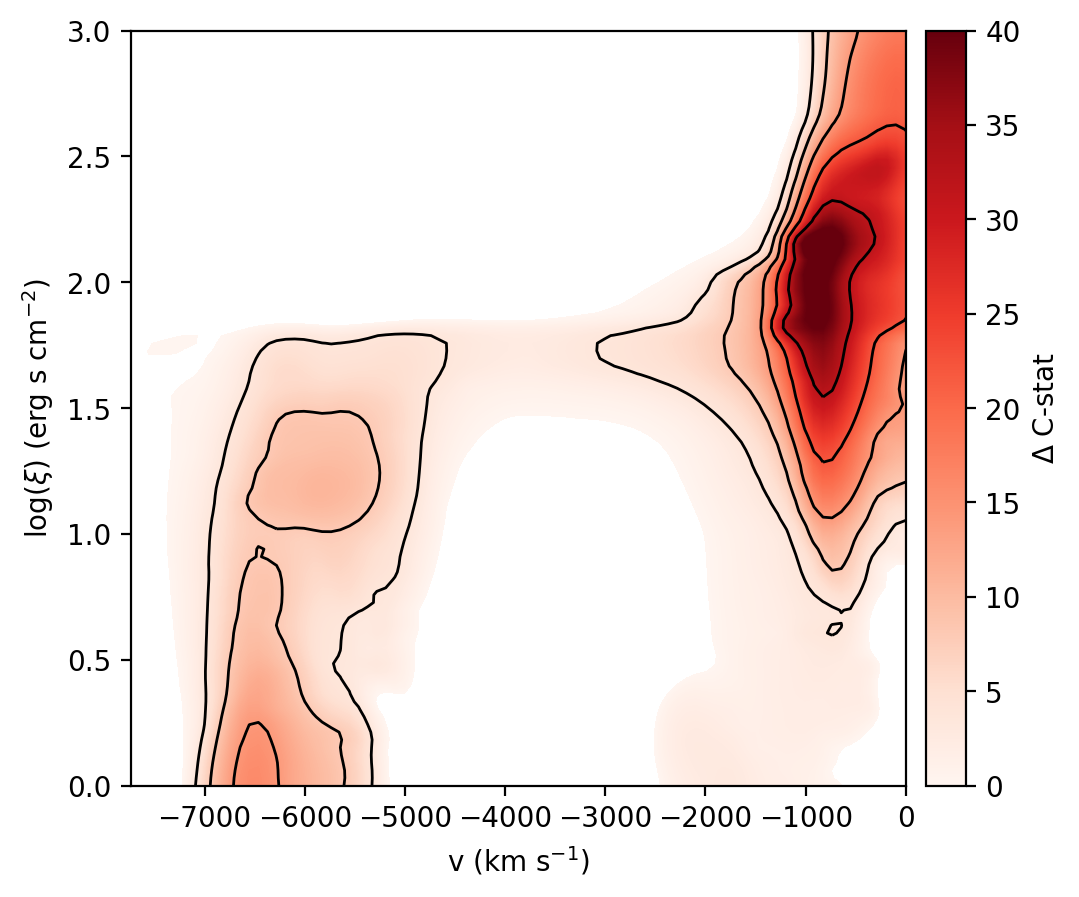}
    \includegraphics[width=0.3\linewidth]{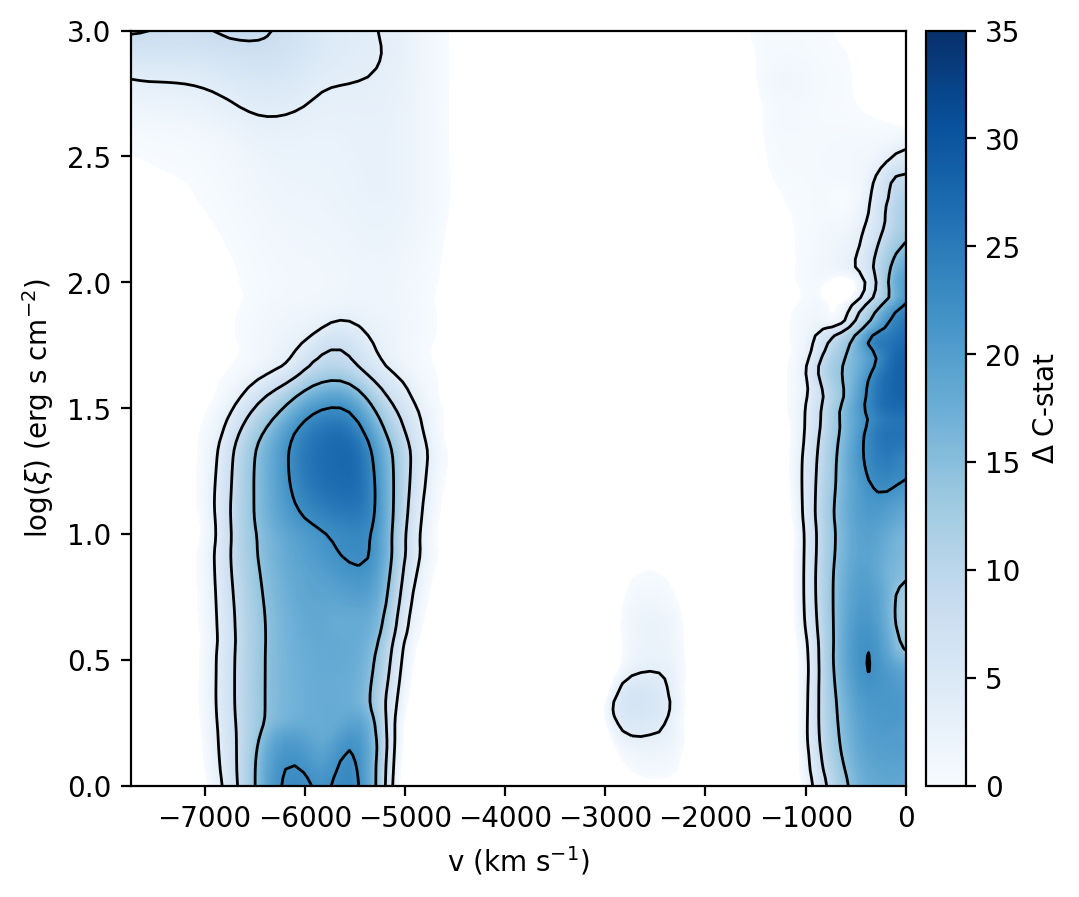}
    \includegraphics[width=0.3\linewidth]{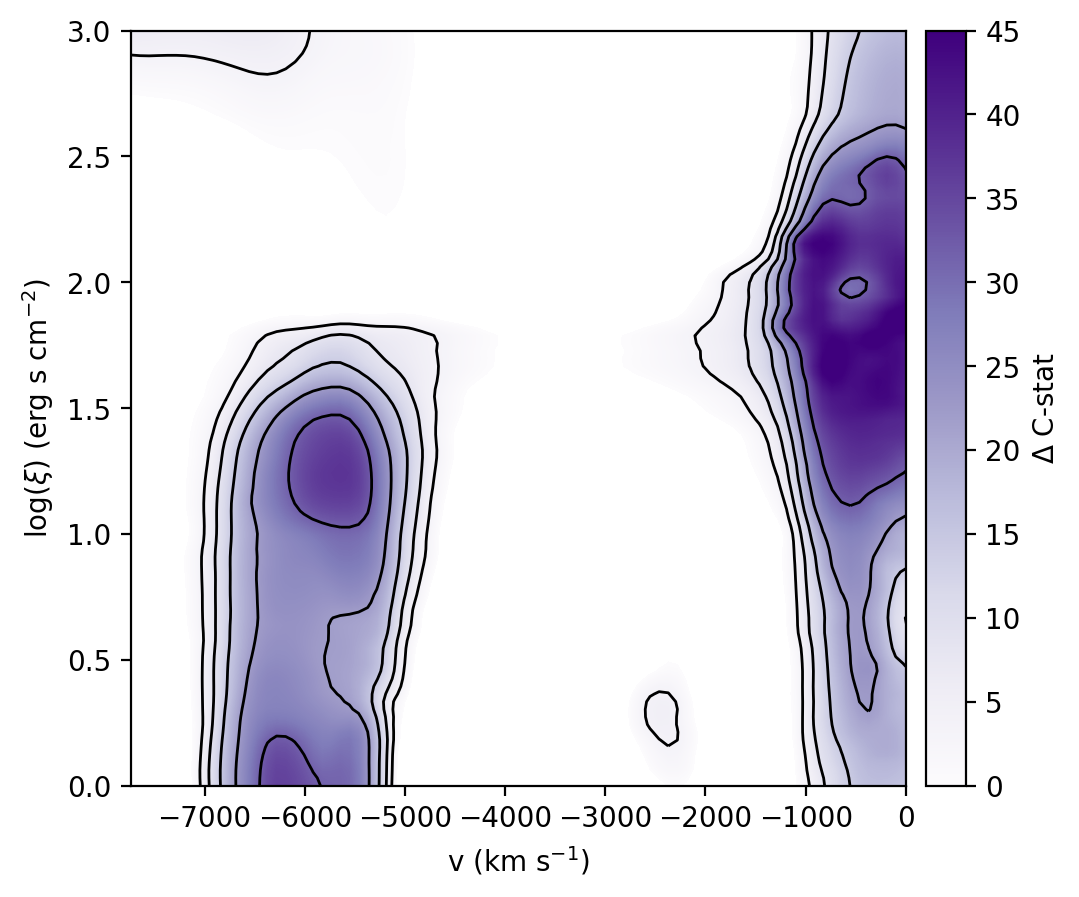}
    \caption{Improvement in fit statistic from adding absorption as a function of ionisation and velocity for the 2018 and 2019 RGS spectra (left and middle) and both combined (right). Black contour lines show the 1--5~$\sigma$ detection thresholds. Two velocity ranges give a significant improvement to the fit, one at $v<1000$~\kms and one at $v\sim6000$~\kms.}
    \label{fig:abs_map}
\end{figure*}

We fit the 2018 and 2019 RGS spectra simultaneously, with two \textsc{pion\_xs} components to model the emission lines (fitting with a single component gives a very poor fit, and misses many of the lines), a power-law continuum, a single layer of warm absorption modelled with \textsc{xstar} (as the \textsc{pion\_xs} model only includes the emission), and Galactic absorption modelled with \textsc{tbnew} \citep{Wilms00}. We also include a \textsc{redge} component to fit the C\textsc{vi} RRC at 26~\AA (observed). We fix the density and velocity of the \textsc{pion\_xs} components between the two spectra, allowing the normalisations and ionisations to vary. Similarly, we fix the photon index of the power-law and the velocity of the warm absorber between the two spectra. For all these parameters, we check that they do not have a major impact on the fit if freed (i.e. they do not give a significant improvement to the fit statistic), and that they are are consistent between the two intervals if freed. The final model, in \textsc{xspec} notation, is therefore: \textsc{tbnew $\times$ xstar $\times$ (powerlaw + pion\_xs + pion\_xs + redge)}. 

We estimate the velocity of the absorption component by scanning the \textsc{xstar} grid across the parameter space, with 100 redshift bins between the source redshift and $z=0$, and 100 ionisation bins between $\log\xi=0$ and $\log\xi=3$ erg~s~cm$^{-2}$. The results of this scan for each spectrum are shown in Fig.~\ref{fig:abs_map}, along with result obtained from fitting both together. Two potential zones of absorption are found, at velocities of $v<1000$~\kms and $v\sim6000$~\kms, both of which give a significant improvement to the fit. Because the continuum level is very low, no absorption lines are visible in the spectrum, so the absorption can only be constrained by its effect on the emission lines. Because the high velocity component is blueshifted and therefore misaligned with the emission spectrum, it does not affect many emission lines. It aligns the \ion{O}{vi} line in absorption with the \ion{O}{vii} resonance emission line in emission, and the \ion{O}{viii} line in absorption with the \ion{O}{vii} 1s-3p line in emission, but no other combinations of strong lines align. This component is therefore a strong candidate for the cause of the observed spectral variability in these lines, and is also consistent with the velocity of absorption previously found in Mrk~335, both in the X-ray and UV bands \citep[e.g.][]{Longinotti19}.

The potential low-velocity absorption, on the other hand, is at a velocity consistent with that found for the emission lines (see below). In this case, many features align, so almost every strong line (with the exception of the forbidden lines and RRC) is modified. Because so many lines are affected, this cannot be the cause of the variability seen between the two spectra, as it would change the strength of many more more features than observed. This means that there is no direct evidence for this absorption. No absorption has previously been identified at these velocities apart from the partial-covering component sometimes used in broad-band fits \citep[e.g.][]{Grupe08, Longinotti19}, which has a column density far too high to be caused by the same material. In our view, the most likely explanation for this improvement in the fit statistic is that adding a low-velocity absorption component modifies the line ratios and widths of the photoionised emission to give a better fit to the observed emission lines. However, there are many other effects that could cause the discrepancy, including assumptions within the \textsc{pion} code, non-equilibrium effects, and non-solar abundances. We are therefore not confident that this represents genuine absorption of the spectrum (although it cannot be ruled out). On this basis, we only include the high velocity component, which we are confident is real, in our fits from this point.

We also run the same test applying the absorption to just the continuum, leaving the photoionised emission unaffected. In this case, no significant detections are returned. Because the continuum contribution to these spectra is so low, the variability in the \ion{O}{vii} triplet must be due to either intrinsic line variability or obscuration of the lines.

The spectra and best-fit model are shown in Fig.~\ref{fig:rgs_spectra}, and the parameters are given in Table~\ref{tab:rgs_pars}. The fit statistic C-stat is 772, for 623 degrees of freedom. The high-ionisation emission component mainly fits the \ion{O}{viii} \Lya\ line and the Fe \textsc{xvii}--\textsc{xx} lines at short wavelengths, while the low ionisation component describes the \ion{O}{vii} triplet and the C, N and Ne lines.

\begin{figure*}
    \centering
    \includegraphics[width=\linewidth]{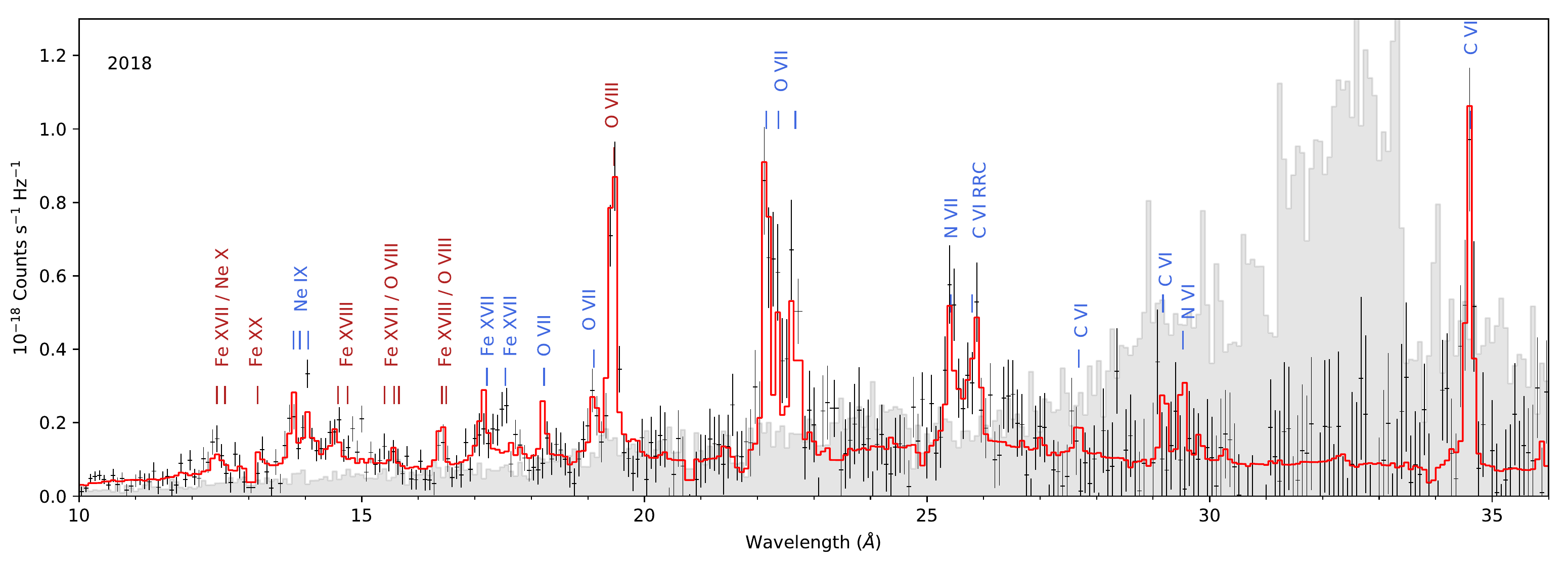}
    \includegraphics[width=\linewidth]{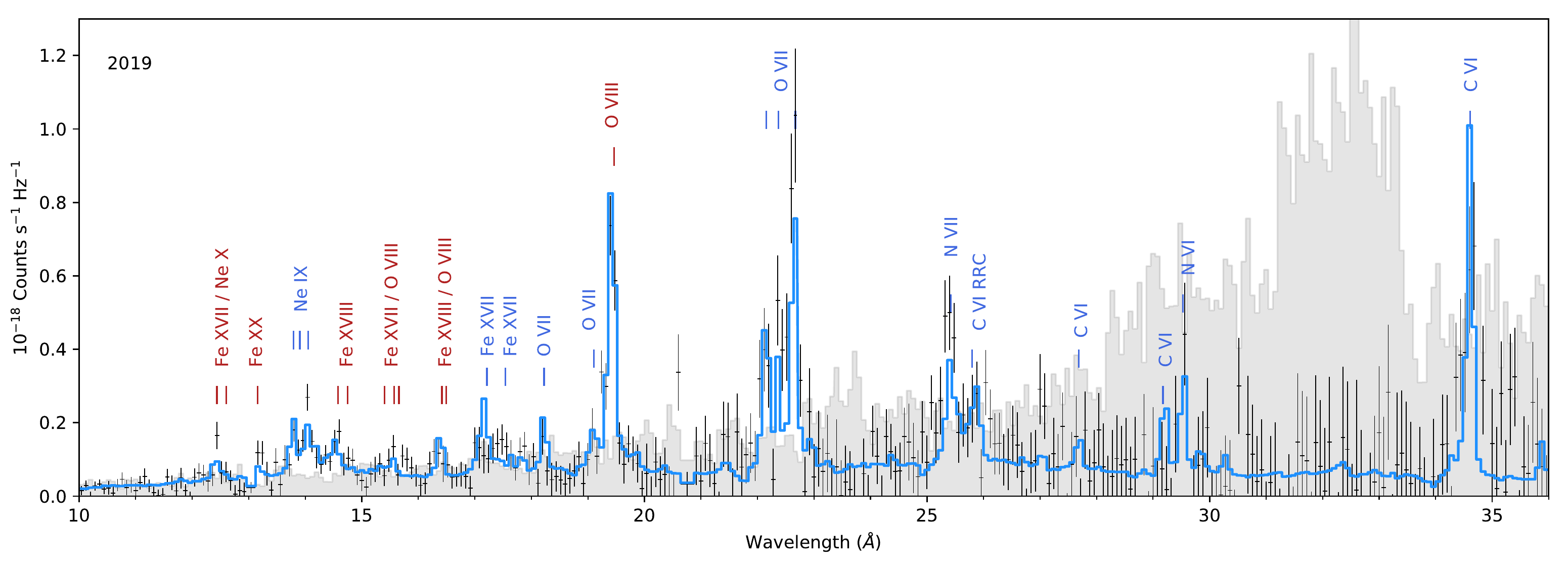}
    \caption{Best-fit RGS spectra for the two \xmm\ observations. The data are fit with a power-law continuum, two photoionised emission components, one warm absorber, and an additional RRC component to fit the C~\textsc{vi} feature. The grey shaded area indicates the background flux, which dominates above $\sim28$~\AA, with the exception of the C\textsc{vi} \Lya\ line. Emission lines produced primarily by the low-ionisation component are labelled in blue, those from the high-ionisation zone in red.}
    \label{fig:rgs_spectra}
\end{figure*}

\begin{table*}
    \centering
    \caption{Best-fit parameters for the joint fit to the two RGS spectra. For parameters that are tied between the two spectra only a single value is given. Errors correspond to the 16 and 84th (one standard deviation) percentiles from the MCMC chains, after marginalising over all other parameters.}
    \begin{tabular}{lcccr}
    \hline
    \hline
Component & Parameter & 2018 & 2019 & Description/Unit\\
\hline
\textsc{tbnew} &   $N_\mathrm{H}$  & \multicolumn{2}{c}{ $3.56\times10^{20*}$ }  &   Column density (cm$^{-2}$)\\
\\
\textsc{xstar} &   $N_\mathrm{H}$  &  $3.2_{-1.0}^{+2.5}\times10^{20}$ & $7.4^{+3.3}_{-1.7}\times10^{20}$  &   Column density (cm$^{-2}$)\\
        
    &   $\log(\xi)$ & $1.45^{+0.06}_{-0.12}$	&	$1.37^{+0.02}_{-0.12}$	& Ionisation parameter (erg~s~cm$^{-2}$)	\\
    &   $v$	&\multicolumn{2}{c}{$-5700^{+200}_{-170}$}	& Velocity (km~s$^{-1}$)	\\
    \\
\textsc{powerlaw}   &  $\Gamma$ &	\multicolumn{2}{c}{$2.39^{+0.14}_{-0.11}$}	& Photon index\\
                    &  Normalisation & $2.3^{+0.1}_{-0.2}\times10^{-4}$ & $1.46\pm0.1\times10^{-4}$ & (photons~keV$^{-1}$~cm$^{-2}$~s$^{-1}$)\\
                    \\
\textsc{pion\_xs} (hot) &   $\log(\xi)$	&	$2.47^{+0.13}_{-0.07}$	& $2.38^{+0.02}_{-0.07}$ & Ionisation parameter (erg~s~cm$^{-2}$)\\
    &   Density & 	\multicolumn{2}{c}{$<9\times10^{10}$}	& (cm$^{-3}$)\\
    &       $v$	&	\multicolumn{2}{c}{$-1260^{+170}_{-100}$}	&	Velocity (km~s$^{-1}$)\\
    & Normalisation &$0.62^{+0.21}_{-0.14}$	&$0.55^{+0.08}_{-0.11}$ & (arbitrary)\\
    \\
    
\textsc{pion\_xs} (cold) &   $\log(\xi)$	&	$1.16^{+0.04}_{-0.04}$	& $1.07^{+0.05}_{-0.12}$ & Ionisation (erg~s~cm$^{-2}$)\\
    &   Density &   \multicolumn{2}{c}{$5.5^{+4.2}_{-3.4}\times10^{10}$} & (cm$^{-3}$)\\
    &       $v$	&	\multicolumn{2}{c}{$0\pm50$}	&	Velocity (km~s$^{-1}$)\\
    & Normalisation &$0.29\pm0.03$	& $0.26^{+0.04}_{-0.03}$& (arbitrary)\\
\hline\hline

    \end{tabular}
    
    $^*$The column density of the Galactic absorption is not well constrained, so we fix it to the literature value \citep{Kalberla05}.
    \label{tab:rgs_pars}
\end{table*}

The best fit model provides a reasonably good description of the data (C-stat/dof=1.24), with only a few features not well fit. The Fe\textsc{xvii} 3s-2p line at 17.5~\AA\ (17.07~\AA\ rest frame) is underestimated by the model, while the 2s-2p at 17.2~\AA\ (16.78~\AA\ rest frame) is slightly overestimated. The N\textsc{vii} \Lya\ line at 25.4\AA\ (24.78\AA\ rest frame) is also underestimated in 2019. The most likely explanation for this is that our model is an oversimplification, and that the emission structure is more complex than two zones with fixed ionisations and densities. If the ionised absorption, modelled with \textsc{xstar}, is not included, then the O triplet is not well fit and the fit statistic worsens by $\Delta$~C-stat of 45, for 3 degrees of freedom.

\begin{figure}
    \centering
    \includegraphics[width=\linewidth]{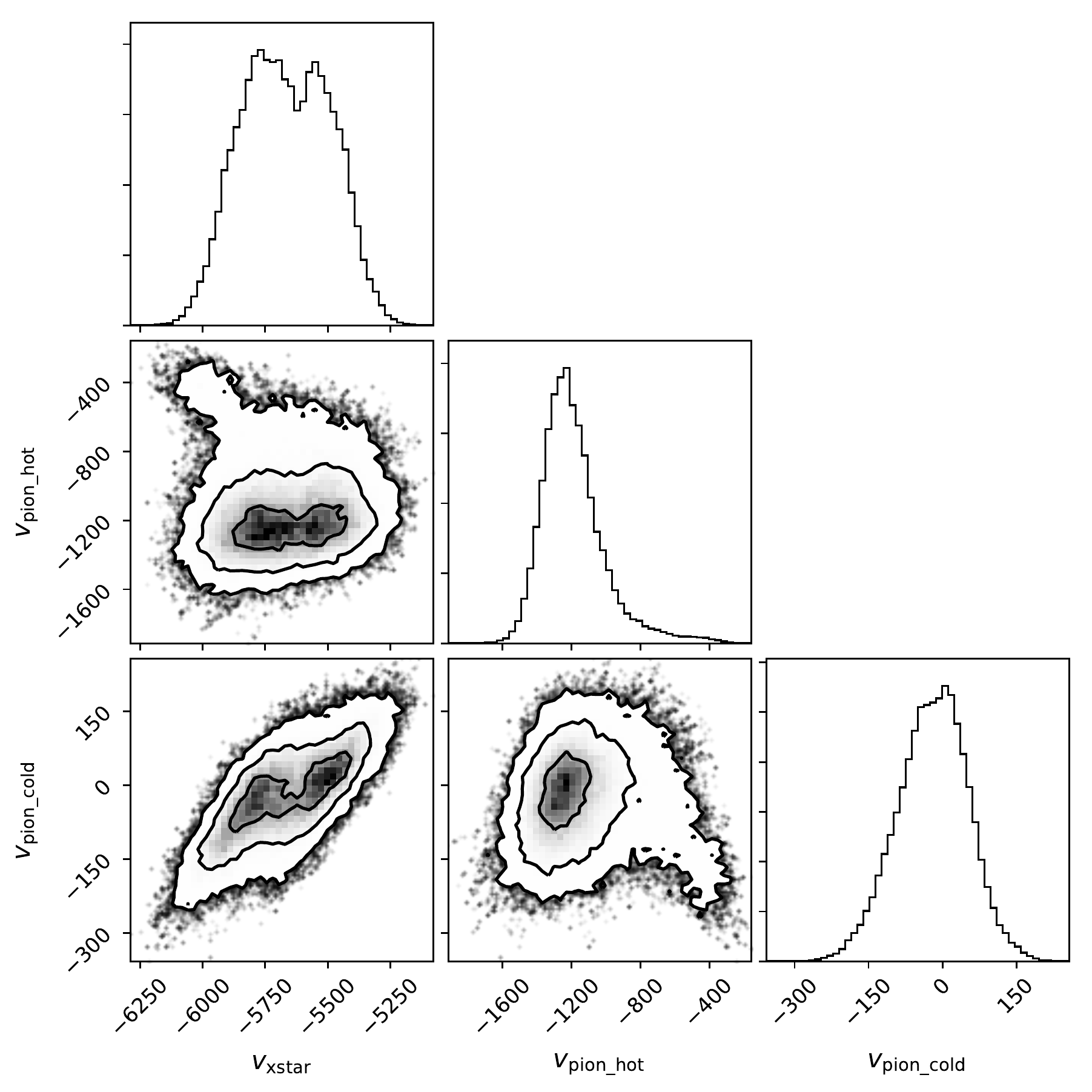}
    \caption{MCMC contours showing the degeneracy between the velocities of the warm absorber and hot and cold photoionised emission components in the RGS spectra. Contours show the 1,2 and 3 sigma levels and shaded regions show the density of points. A strong degeneracy is visible between the velocities of the warm absorber and the cold photoionised component.}
    \label{fig:z_contours}
\end{figure}

The changes between the two spectra are consistent with the effect of a drop in the ionising flux, with the drop in absorption ionisation driving most of the variability. All three ionisation parameters drop by a similar amount ($\Delta\log\sim-0.09$, a $\sim20\%$ drop) between 2018 and 2019. This is smaller than the drop in continuum flux, but consistent with the drop in overall flux when all emission components are considered (see Section~\ref{sec_broadband}), implying that both emitters and the absorber are seeing a similar radiation environment, and that a large fraction of the ionising flux is reprocessed, so the ionisation of these components does not have to perfectly track the continuum. 

The differences between the spectra are largely attributable to the changing absorption. In particular, the change in the ratios of the O\textsc{vii} triplet lines is caused by increased absorption of the resonance line by \ion{O}{vi} in the warm absorber, either due to the increase in column density or drop in ionisation of the warm absorber.

\begin{figure}
    \centering
    \includegraphics[width=0.9\linewidth]{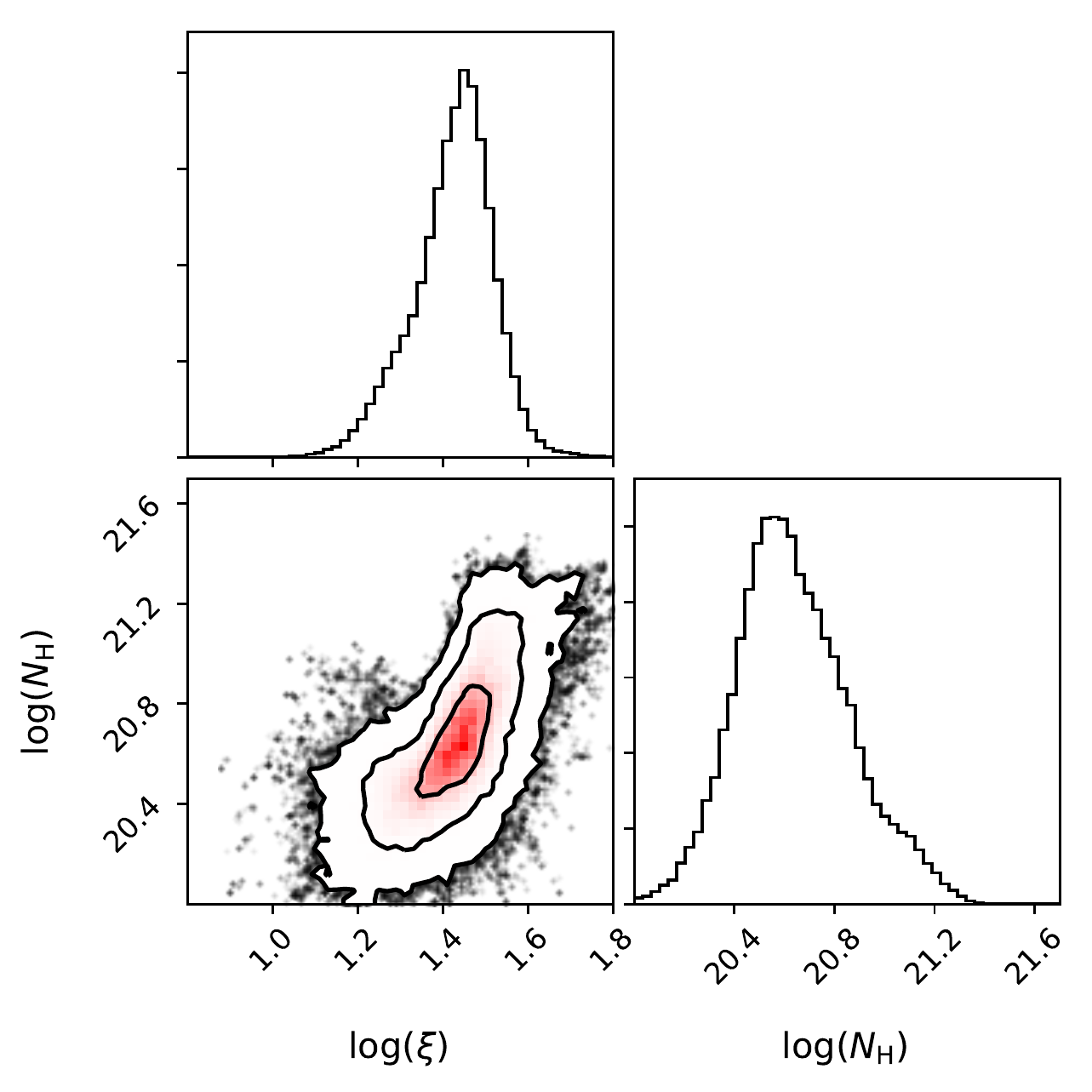}
    \includegraphics[width=0.9\linewidth]{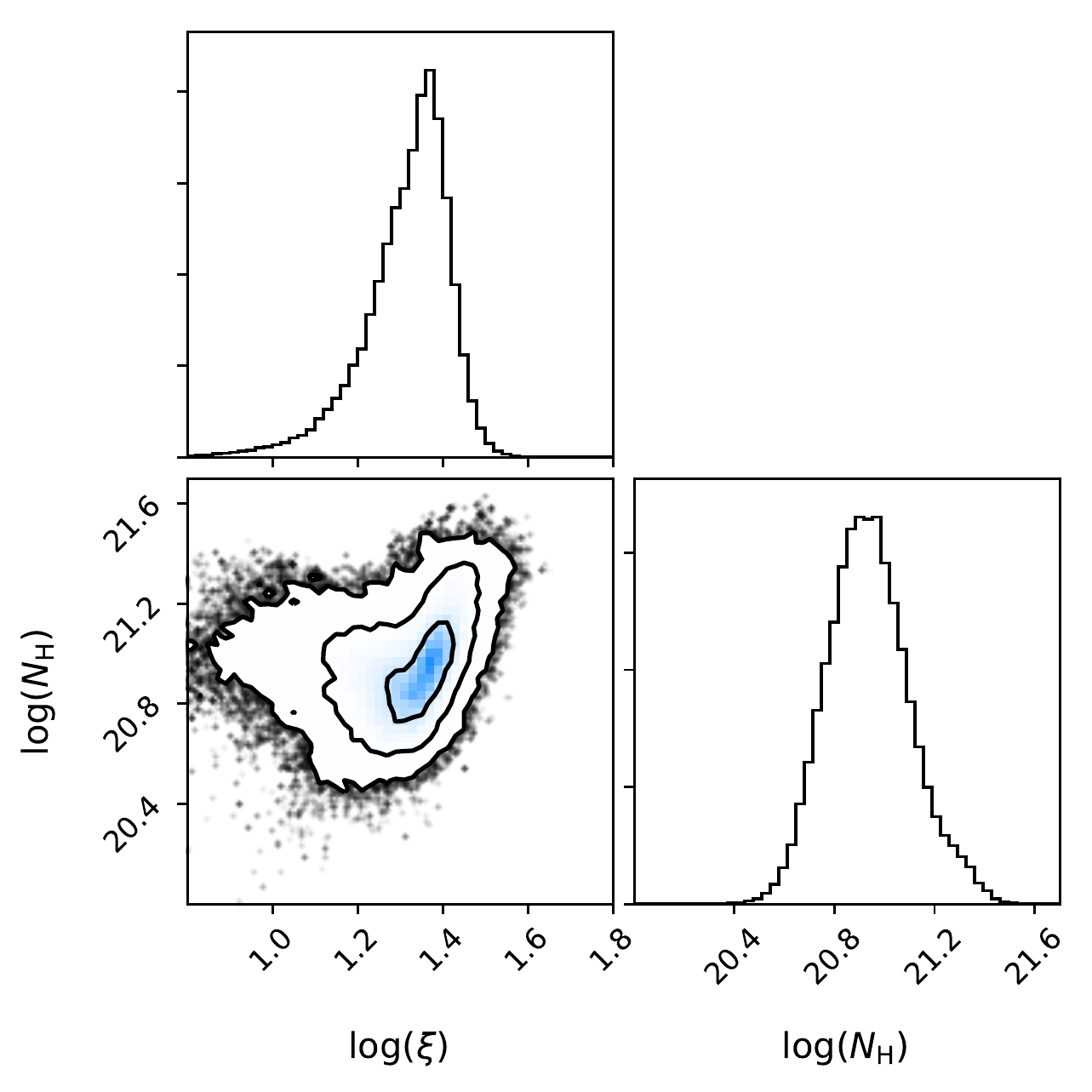}
    \caption{MCMC contours showing the degeneracy between ionisation (in erg~s~cm$^{-2}$) against column density (cm$^{-2}$) for the 2018 (top) and 2019 (bottom) RGS spectra. Contours show the 1,2 and 3 sigma levels and shaded regions show the density of points.}
    \label{fig:wa_degeneracies}
\end{figure}

We search for degeneracies using the \textsc{xspec\_emcee}\footnote{\url{https://github.com/jeremysanders/xspec_emcee}} implementation of the \textsc{emcee} code \citep{Foreman-Mackey13}. Most parameters are independent, with two exceptions. Firstly, the redshift of the absorber is strongly correlated with that of the colder photoionised emission (Fig.~\ref{fig:z_contours})\footnote{All contour figures were generated using the \textsc{corner} code \citep{corner}.}, as required by the need to absorb a specific emission line. Indeed, this degeneracy is the main source of uncertainty in the velocity of the warm absorber: if the velocity of the cold emission component is fixed to zero, the measurement of the absorber velocity is $-5600_{-50}^{+100}$~km~s$^{-1}$. The second degeneracy is between the ionisation and column density of the absorber, in both 2018 and 2019 (Fig.~\ref{fig:wa_degeneracies}). This is due to the fact that we are effectively fitting only a very small number of absorption lines, where they coincide with photoionised emission lines, so the main effect of both column density and ionisation is to change the strength of the absorption, with minimal broader effect. In particular, with a typical absorbed continuum, the ionisation would usually be much more strongly constrained. We note that the contours shown in Fig.~\ref{fig:wa_degeneracies} do not overlap out to $\gtrsim2\sigma$, so variability of the warm absorber is required to fit the data. This is not the case for the ionised emission, where the ionisation parameters and normalisations are consistent with being constant between the two spectra at the $1\sigma$ level.

Finally, we use a small test \textsc{pion\_xs} grid with the parameters fixed to the best fit values given in Table~\ref{tab:rgs_pars}, but with the RMS velocity of the gas added as a free parameter to measure the velocity broadening of the photoionised emission lines. This improves the fit by a $\Delta$~C-stat of $\sim20$, for two degrees of freedom. The best-fit RMS velocities for the hot and cold emission components are $1000\pm200$~km~s$^{-1}$ and $380\pm80$~km~s$^{-1}$, respectively.

\subsection{Broad-band fitting}
\label{sec_broadband}

\begin{figure*}
    \centering
    \includegraphics[width=0.8\linewidth]{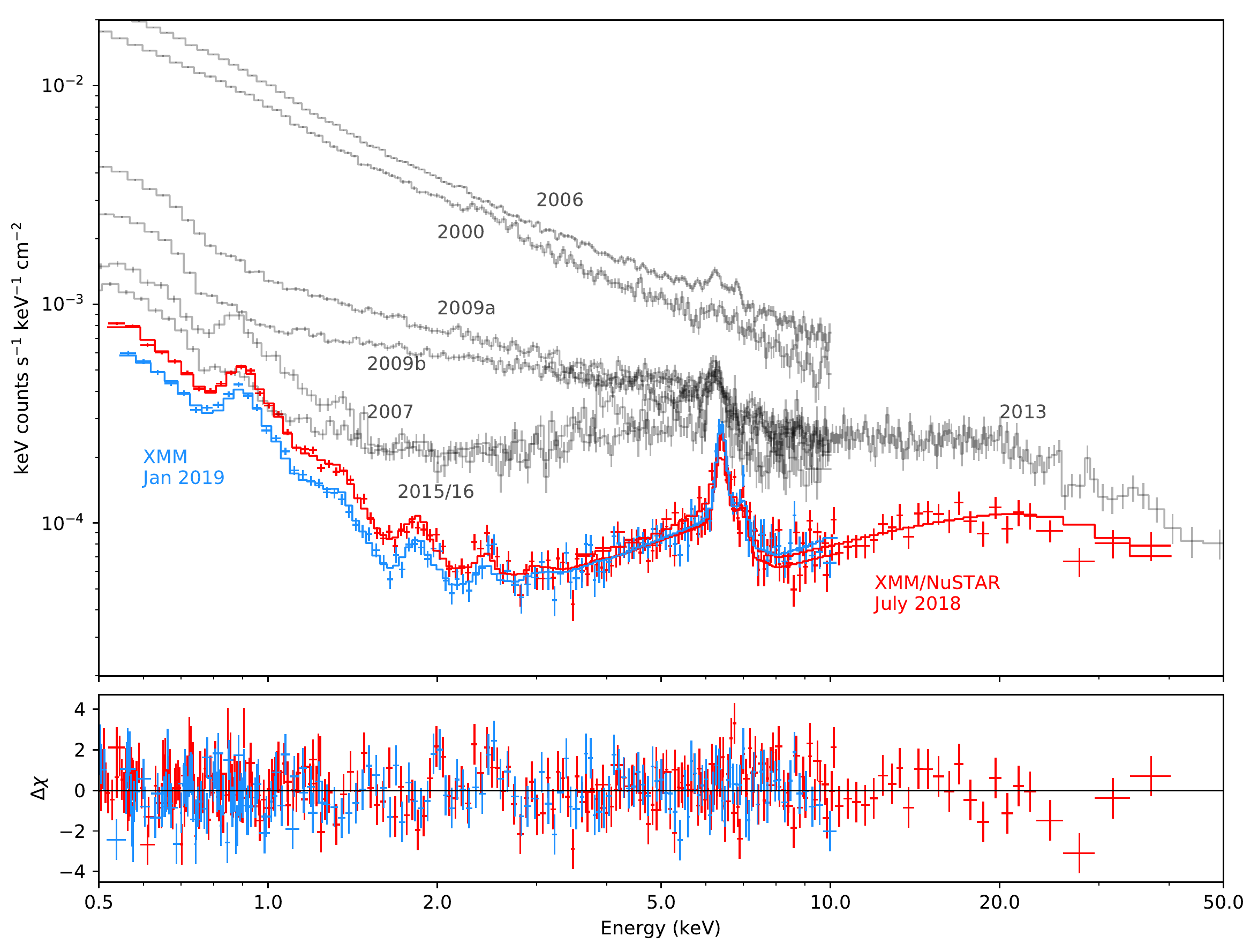}
    \caption{Broad-band best fit to the 2018 and 2019 spectra (top) and residuals (bottom). RGS spectra are not shown in the main panel for clarity, but are included in the fit and shown in the residuals panel. The spectra are corrected for the effective area of the detectors, but not unfolded from the instrumental response, so any spectral features are genuinely present in the data.}
    \label{fig:broadband}
\end{figure*}

We now extend our fitting to the full broad-band spectrum of \src , again fitting both 2018 and 2019 data simultaneously, and including all instruments. These spectra are shown in Fig.~\ref{fig:broadband}, along with previous \xmm\ and \nustar\ spectra. The spectra are very similar, with a small drop in flux ($\sim20$\%) at low energies. Above 3~keV they are consistent with each other. The broad-band spectra are a factor of $\sim5$ lower in flux than the average in 2017--18.

In addition to the photoionised emission and absorption and power-law continuum we used as the model for the RGS spectra, for the broad-band fit we include a black body for the soft excess, and a distant reflection component \citep[modelled with \textsc{xillver}:][]{Garcia13}.
The spectrum of Mrk~335 frequently displays a broad excess around 6--7~keV, which has been interpreted as evidence of either partial-covering absorption or relativistic reflection. In this case, the continuum level is so low that we have little prospect for distinguishing the two statistically. We therefore consider both cases. Firstly, we fit with the relativistic reflection model \textsc{relxill} \citep{Garcia14}, then we use the partial covering model \textsc{zpcfabs}, applied to the power-law and black body continuum. The two models, in \textsc{xspec} notation, are therefore:
\begin{enumerate}
    \item \textsc{tbnew $\times$ (xstar $\times$ (powerlaw + bb + relxill + pion\_xs + pion\_xs) + xillver)}
    \item \textsc{tbnew $\times$ (xstar $\times$ ( zpcfabs $\times$ (powerlaw + bb) + pion\_xs + pion\_xs) + xillver)}
\end{enumerate}

\begin{figure*}
    \centering
    \includegraphics[width=0.8\linewidth]{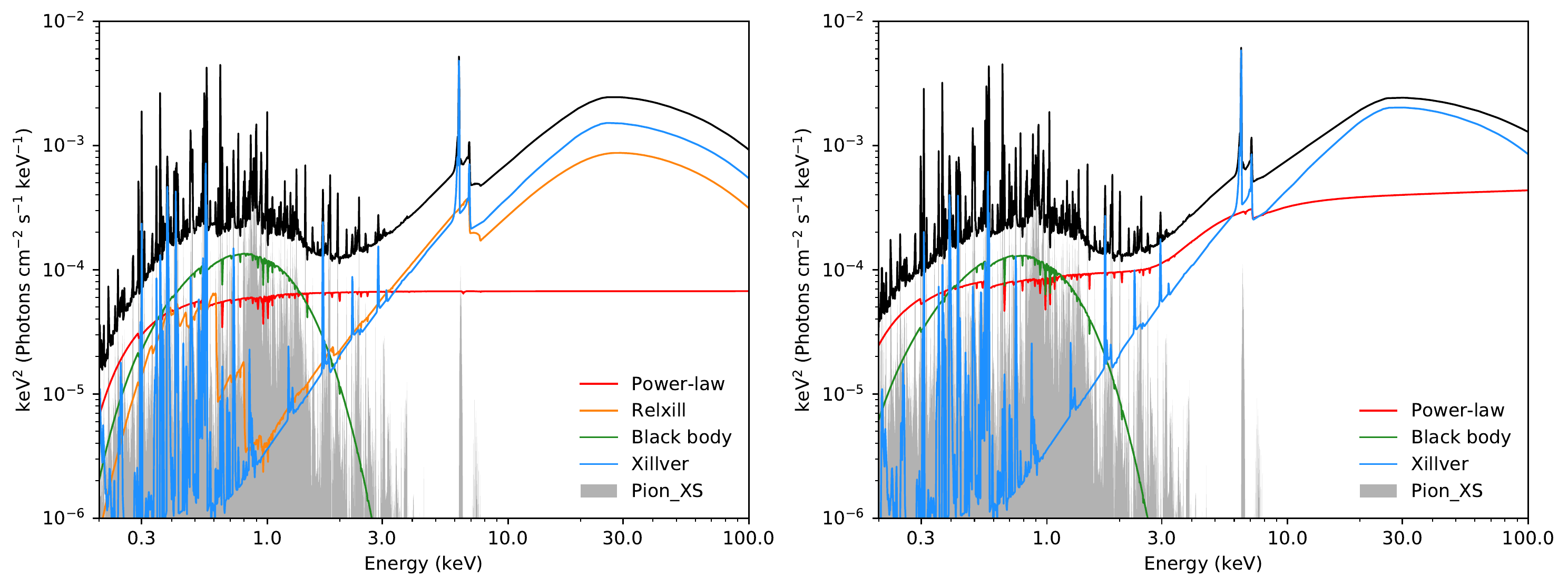}
    \caption{Relativistic reflection (left) and partial covering (right) models for the broad-band spectrum shown in Fig.~\ref{fig:broadband}.Both models give an equivalent fit to the data, and the primary purpose of the additional component in each case is to fit the excess flux around 7~keV that is not accounted for by distant reflection.}
    \label{fig:broadbandmodels}
\end{figure*}

The two models give comparable fit statistics (in both cases, the minimum $\chi^2$ found by the MCMC chains was 1277, with 969 degrees of freedom for the  partial covering model and 970 for the reflection model). The fit with the reflection model (we show only one model as the two are indistinguishable by eye) is shown in Fig.~\ref{fig:broadband}, and both models are shown in Fig.~\ref{fig:broadbandmodels}. 

The phenomenological black body component is required to fit the soft excess in both cases, removing it worsens the fit by $\Delta\chi^2\sim200$, for two degrees of freedom. There are several possible explanations for this component. It could be a result of only including a single layer of warm absorption, when the real absorption structure is more complex \citep[as seen previously by][]{Longinotti13}, or it could be that a more sophisticated reflection model, taking into account the disk density, is required, as these models can give a stronger soft excess \citep[][Jiang et al., submitted]{Garcia16}. Contamination from X-ray binaries or ultra-luminous X-ray sources is unlikely as the origin of this component, as the luminosity is high ($5\times10^{41}$~erg~s$^{-1}$), indicating a likely nuclear origin. We do not explore these models here, as the soft band is dominated by the photoionised emission lines and our ability to constrain complex reflection or absorption models with this data is very limited.

\begin{table*}
    \centering
    \caption{Best-fit parameters for the joint fit to broadband spectra with the \textsc{relxill} model. For parameters that are tied between the two spectra only a single value is given.}
    \begin{tabular}{lcccr}
    \hline
    \hline
Component & Parameter & 2018 & 2019 & Description/Unit\\
\hline
\textsc{tbnew} &   $N_\mathrm{H}$  & \multicolumn{2}{c}{ $3.56\times10^{20*}$ }  &   Column density (cm$^{-2}$)\\
\\
\textsc{xstar} &   $N_\mathrm{H}$  &  $8.2_{-0.2}^{+0.4}\times10^{20}$ & $8.8^{+0.3}_{-0.2}\times10^{20}$  &   Column density (cm$^{-2}$)\\
        
    &   $\log(\xi)$ & $1.47\pm0.05$	&	$1.36^{+0.03}_{-0.10}$	& Ionisation (erg~s~cm$^{-2}$)	\\
    &   $v$	&\multicolumn{2}{c}{$-6089\pm63$}	& Velocity (km~s$^{-1}$)	\\
    \\
\textsc{powerlaw}   &  $\Gamma$ &	$2.000\pm0.007$& $1.88^{+0.041}_{-0.049}$	& Photon index\\
                    &  Normalisation & $6.7^{+0.5}_{-0.4}\times10^{-5}$ &  $5.7^{+0.4}_{-0.7}\times10^{-5}$& (photons~keV$^{-1}$~cm$^{-2}$~s$^{-1}$)\\
                    \\
\textsc{black body}   &  kT & $0.185\pm0.005$	& $0.174^{+0.005}_{-0.007}$	& Temperature (keV)\\
                    &  Normalisation &  $4.8^{+0.3}_{-0.4}\times10^{-6}$	& $4.1^{+0.5}_{-0.3}\times10^{-6}$ & ($10^{37}$~erg~s$^{-1}$~kpc$^{-2}$)\\
                    \\
\textsc{pion\_xs} (hot) &   $\log(\xi)$	&	$2.34^{+0.01}_{-0.03}$	& $2.29\pm0.023$ & Ionisation parameter (erg~s~cm$^{-2}$)\\
    &   Density & 	\multicolumn{2}{c}{$<5\times10^{9}$}	& (cm$^{-3}$)\\
    &       $v$	&	\multicolumn{2}{c}{$-386\pm51$}	&	Velocity (km~s$^{-1}$)\\
    & Normalisation & $0.95\pm0.06$	& $0.92\pm0.06$ & (arbitrary)\\
    \\

\textsc{pion\_xs} (cold) &   $\log(\xi)$	&	$0.83^{+0.06}_{-0.05}$	& $0.85^{+0.05}_{-0.08}$ & Ionisation parameter (erg~s~cm$^{-2}$)\\
    &   Density &   \multicolumn{2}{c}{$<1.3\times10^{10}$} & (cm$^{-3}$)\\
    &       $v$	&	\multicolumn{2}{c}{$-236\pm66$}	&	Velocity (km~s$^{-1}$)\\
    & Normalisation &$0.24\pm0.02$	& $0.24^{+0.02}_{-0.03}$& (arbitrary)\\
\\
    
\textsc{xillver}    &       $A_\mathrm{Fe}$	&	\multicolumn{2}{c}{$1^*$}	&	Iron abundance (solar)\\
    &   $\log(\xi)$	&	\multicolumn{2}{c}{$0^*$} & Ionisation (erg~s~cm$^{-2}$)\\
    &   $E_\mathrm{cut}$	&	\multicolumn{2}{c}{$300^*$} & High-energy cutoff (keV)\\
    &  Normalisation & $2.7^{+0.1}_{-0.2}\times10^{-5}$ & $2.9\pm0.2\times10^{-5}$ & (arbitrary)\\
                    \\

\textsc{relxill}   &       $A_\mathrm{Fe}$	&	\multicolumn{2}{c}{$2.1\pm0.3$}	&	Iron abundance (solar)\\
    &   $\log(\xi)$	&	\multicolumn{2}{c}{$<0.3$} & Ionisation parameter (erg~s~cm$^{-2}$)\\
    &   $i$	&	\multicolumn{2}{c}{$50^*$} & Inclination (degrees)\\
    &   $q$	&	\multicolumn{2}{c}{$3^*$} & Emissivity index\\
    &   $a$	&	\multicolumn{2}{c}{$0.98^*$} & Spin\\
    &  Normalisation & $2.7^{+0.1}_{-0.2}\times10^{-5}$ & $2.9\pm0.2\times10^{-5}$ & (arbitrary)\\
                    \\
                    
\textsc{constant}& RGS/EPIC-pn  & \multicolumn{2}{c}{$0.87\pm0.02$}	& Constant scaling factor\\
                 & FPM/EPIC-pn  & \multicolumn{2}{c}{$1.26\pm0.06$}	& Constant scaling factor\\

\hline\hline

    \end{tabular}
    
    $^*$The column density of the Galactic absorption is not well constrained, so we fix it to the literature value \citep{Kalberla05}. We also fix most of the reflection parameters to values representative of the results of previous authors for Mrk~335, as they are not well constrained by this spectrum. The one exception is the iron abundance of the relativistic reflector, which we leave free to give a parameter to account for the relative strength of the iron line and Compton hump.
    \label{tab:braodband_relxill_pars}
\end{table*}

\begin{table*}
    \centering
    \caption{Best-fit parameters for the joint fit to broadband spectra with the \textsc{zpcfabs} model. For parameters that are tied between the two spectra only a single value is given.}
    \begin{tabular}{lcccr}
    \hline
    \hline
Component & Parameter & 2018 & 2019 & Description/Unit\\
\hline
\textsc{tbnew} &   $N_\mathrm{H}$  & \multicolumn{2}{c}{ $3.56\times10^{20*}$ }  &   Column density (cm$^{-2}$)\\
\\
\textsc{xstar} &   $N_\mathrm{H}$  &  $1.1\pm0.3\times10^{20}$ & $1.1^{+0.2}_{-0.3}\times10^{20}$  &   Column density (cm$^{-2}$)\\
        
    &   $\log(\xi)$ & $1.50\pm0.05$	&	$1.38^{+0.03}_{-0.08}$	& Ionisation (erg~s~cm$^{-2}$)	\\
    &   $v$	&\multicolumn{2}{c}{$-6097\pm72$}	& Velocity (km~s$^{-1}$)	\\
    \\
\textsc{zpcfabs} &   $N_\mathrm{H}$  &  $2.47\pm0.01\times10^{22}$ & $1.61\pm0.03\times10^{22}$  &   Column density (cm$^{-2}$)\\
                &   $f_\mathrm{cov}$    &  $0.74\pm0.02$ & $0.72\pm0.03$  & Covering fraction\\
        
    \\
\textsc{powerlaw}   &  $\Gamma$ &	$1.95\pm0.03$& $1.83^{+0.03}_{-0.12}$	& Photon index\\
                    &  Normalisation & $3.7\pm0.3\times10^{-4}$ &  $3.0\pm0.4\times10^{-4}$& (photons~keV$^{-1}$~cm$^{-2}$~s$^{-1}$)\\
                    \\
\textsc{black body}   &  kT & $0.170\pm0.005$	& $0.157\pm0.006$	& Temperature (keV)\\
                    &  Normalisation &  $1.8\pm0.2\times10^{-5}$	& $1.6\pm0.2\times10^{-5}$ & ($10^{37}$~erg~s$^{-1}$~kpc$^{-2}$)\\
                    \\
\textsc{pion\_xs} (hot) &   $\log(\xi)$	&	$2.34\pm0.02$	& $2.30\pm0.023$ & Ionisation parameter (erg~s~cm$^{-2}$)\\
    &   Density & 	\multicolumn{2}{c}{$<1.5\times10^{10}$}	& (cm$^{-3}$)\\
    &       $v$	&	\multicolumn{2}{c}{$-355^{+45}_{-69}$}	&	Velocity (km~s$^{-1}$)\\
    & Normalisation & $0.96\pm0.06$	& $0.93\pm0.07$ & (arbitrary)\\
    \\

\textsc{pion\_xs} (cold) &   $\log(\xi)$	&	$0.82^{+0.05}_{-0.07}$	& $0.80^{+0.06}_{-0.08}$ & Ionisation parameter (erg~s~cm$^{-2}$)\\
    &   Density &   \multicolumn{2}{c}{$<1.5\times10^{10}$} & (cm$^{-3}$)\\
    &       $v$	&	\multicolumn{2}{c}{$-265\pm91$}	&	Velocity (km~s$^{-1}$)\\
    & Normalisation &$0.25\pm0.02$	& $0.24\pm0.02$& (arbitrary)\\
\\
    
\textsc{xillver}    &       $A_\mathrm{Fe}$	&	\multicolumn{2}{c}{$1^*$}	&	Iron abundance (solar)\\
    &   $\log(\xi)$	&	\multicolumn{2}{c}{$0^*$} & Ionisation (erg~s~cm$^{-2}$)\\
    &   $E_\mathrm{cut}$	&	\multicolumn{2}{c}{$300^*$} & High-energy cutoff (keV)\\
    &  Normalisation & $3.0^{+0.2}_{-0.1}\times10^{-5}$ & $3.2\pm0.2\times10^{-5}$ & (arbitrary)\\
                    \\

\textsc{constant}& RGS/EPIC-pn  & \multicolumn{2}{c}{$0.86\pm0.02$}	& Constant scaling factor\\
                 & FPM/EPIC-pn  & \multicolumn{2}{c}{$1.35\pm0.06$}	& Constant scaling factor\\

\hline\hline

    \end{tabular}
    
    $^*$The column density of the Galactic absorption is not well constrained, so we fix it to the literature value \citep{Kalberla05}.
    \label{tab:braodband_zpcfabs_pars}
\end{table*}

% \section{8~keV feature}
% \label{sec_ufo}

We also test the inclusion of an additional ionised layer of partial-covering warm absorption in both reflection and absorption models. This is motivated by the additional absorption seen in the HST spectrum, which may be associated with the mildly ionised partial covering absorption required to fit the broadband X-ray spectrum in \citet{Longinotti19}. We use the \textsc{zxipcf} model \citep{Reeves08}, applied to all emission components apart from the distant reflection. In both reflection and partial covering dominated models, this extra component only improves the fit by $\Delta\chi^2<5$ in both cases, for three additional degrees of freedom, and we cannot meaningfully constrain any of the parameters. However, it is still entirely possible that this component is present, as there should be additional X-ray absorption associated with the UV absorption.

The contribution of the distant reflection to the soft lines is relatively small. We test adding this \textsc{xillver} component back in to the RGS fit, with the parameters fixed at the values from the broad-band fit we use the \textsc{relxill} fit to the RGS, EPIC-pn and \nustar\ data presented in Table~\ref{tab:braodband_relxill_pars}, to see what impact it has on the parameters of the photoionised emission. The effect of adding this component back in to the RGS fit is negligible, because it mainly contributes a small amount to the C and N lines at large wavelengths where there is very little signal in the RGS. For the main O lines, and the lower wavelength lines where the RGS spectrum is well constrained, the distant reflection contributes less than $10\%$ of the flux.

In the 2018 \xmm\ spectra, an absorption line like feature is visible at $\sim8.5$~keV.
These features are usually interpreted as evidence for absorption from so-called ultra-fast outflows (UFOs), winds with velocities of $\sim0.03-0.3c$. Indeed, \citep{Gallo19} found tentative evidence for such blueshifted absorption in Mrk~335, with a velocity of $\sim0.12c$. While some such detections are confirmed at huge statistical significance \citep[e.g.][]{Nardini15}, there is a large grey area of marginal significance detections, and the field is plagued by false detections \citep{Vaughan08}. In this case we have reason to be cautious, as the signal-to-noise is relatively low and the continuum is dominated by distant reflection from the torus, so any wind covering it would have to be extremely large and unlikely to be so fast or highly ionised.

A clear feature is present in both the EPIC-pn and EPIC-MOS in 2018. The \nustar\ data do not show a corresponding feature, but this is not conclusive as the resolution of \nustar\ is not optimal for the detection of absorption lines. If we fit the 3--10~keV spectrum of the EPIC-pn, EPIC-MOS and \nustar\ simultaneously with neutral reflection alone, the improvement in fit statistic for adding a Gaussian absorption line at 8.3~keV (observed frame) is $\Delta\chi^2=12$, for two degrees of freedom. However, if we use a more complex continuum (either by adding relativistic reflection with \textsc{relxill} or partial covering with \textsc{zpcfabs}) the improvement in fit statistic for an additional absorption line drops to $\sim6$, which is clearly not significant. 
Finally, we note that no such feature is present in the 2019 spectrum for either the pn or MOS, so we conclude that this is not a real absorption feature and is most likely just a statistical fluctuation.

%%%%%%%%%%%%%%%%%%%%%%%%%%%%%%%%%%%%%%%%%%%%%%%%%%

\section{Discussion}

\begin{figure}
    \centering
    \includegraphics[width=\linewidth]{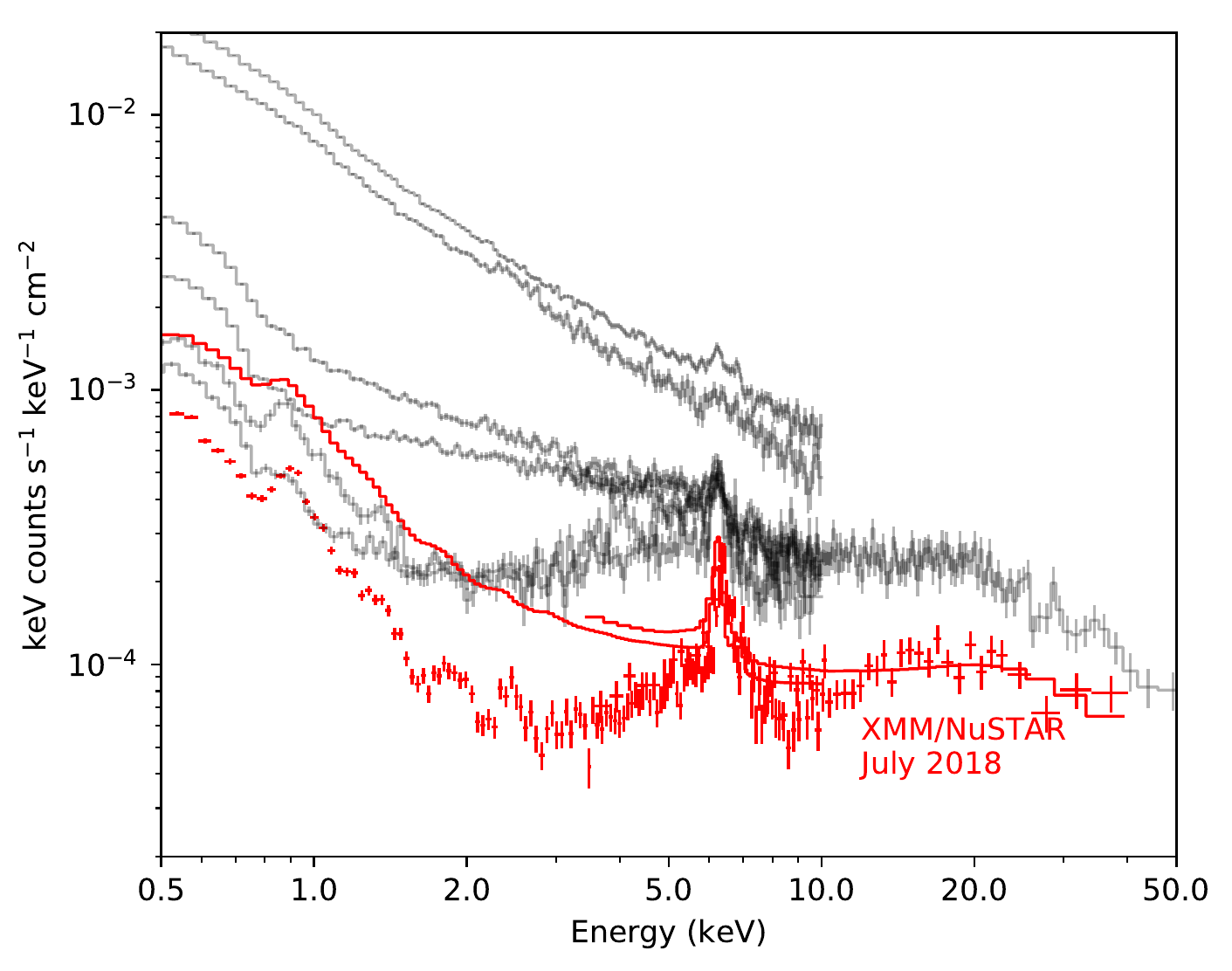}
    \caption{Broad-band spectra of Mrk 335, showing the partial covering model from the 2018 spectrum with the absorption removed. The red points and lines show the \xmm\ and \nustar\ spectra, and corresponding models, in 2018. The grey points show the spectra from previous observations, as in Fig.~\ref{fig:broadband}.}
    \label{fig:pcmodel_noabs}
\end{figure}

% \red{Nature of flux drop}
It is clear that the majority of the flux drop in Mrk~335 must be intrinsic to the source, regardless of the model used for the X-ray spectra. The flux drop is observed in both the X-rays and UV at a similar level (a factor of 3--5 below the 2017/18 average), and the drop in X-ray flux extends throughout the band, including the high energy \nustar\ data, unlike flux drops caused by obscuration \citep[e.g.][]{Parker14_mrk1048, Kaastra14, Mehdipour17}. This does not rule out the presence of any obscuration by cold gas in the spectrum however, as we discuss in section~\ref{sec_broadband}. The broad-band \xmm/\nustar X-ray spectrum is consistent with either partial covering absorption or relativistic reflection (or both), and the presence of complex absorption in Mrk~335 is well established. If we take the partial covering model and remove the absorption, as shown in Fig.~\ref{fig:pcmodel_noabs}, the model does not approximate the spectrum of the source in other intervals: the flux is too great at low energies, and non-existent at high energies. Therefore, even if there is a significant column of partial covering absorption in \src , the current low flux interval must be intrinsic and not caused by an increase in absorbing column or covering fraction.

In theory, a partial-covering absorber that is completely opaque could produce such a broad-band drop in flux, both in the UV and X-ray bands. However, such high density gas would in practise be embedded in less dense, but still Compton-thick material, which would likely have to fully cover the X-ray source in order for the UV to be sufficiently absorbed. This would leave clear signatures of the absorption in both the X-ray and UV bands, so we can safely discount this extreme partial-covering scenario.

The broad-band X-ray spectrum of Mrk~335 (section~\ref{sec_broadband}) is dominated by soft emission lines at low energies and distant reflection at high energies. Since the flux drop is intrinsic to the source, the dominance of this reprocessed emission must be due to the light travel time. The regions responsible for this emission must be located far enough from the X-ray source that they have not yet seen the 2018 drop in flux, and thus have not yet responded to it. We explore this further when we discuss the location of the emission and absorption regions.

These spectra are similar to those found in NGC~5548 and NGC~3783 \citep{Kaastra14,Mao18,Mao19_ngc3783}, where major obscuration events caused by outflowing absorption cause large flux drops at low energies and reveal the soft X-ray emission lines. In this case, however, the flux drop is intrinsic to the source, and we see the effect of the outflow acting on the emission lines. Based on this result, it may be productive to search for similar variability of emission lines driven by wind variability in other sources, although this may require the large flux changes seen in Mrk~335 to modify the ionisation of the wind sufficiently.

In the past, the Fe~K band of Mrk~335 has been fit with both relativistic reflection and partial covering models. The initial work on the first low state by \citet{Grupe07,Grupe08} found that it was not possible to distinguish the two models with the available spectra. Arguably the strongest evidence for the presence of relativistic reflection is the detection of an Fe~K lag \citep{Kara13}, while the presence of multiple layers of complex absorption in both X-rays and UV is well established \citep{Longinotti13,Longinotti19}.
\citet{Gallo15} compared both models, and found that relativistic reflection was more likely to be the dominant cause of the excess Fe~K flux, however \citet{Longinotti19} found that the inclusion of both a reflection component and a ionised partial covering absorber was required to fit the 2015/16 \xmm\ spectrum. In the spectra presented here, either model gives an acceptable fit to the data, as the continuum provides only a very small amount of the total flux and cannot be well constrained. A slight additional curvature in the Fe band is required in the broad-band RGS/EPIC-pn/\nustar model (section~\ref{sec_broadband}) to adequately fit the data, but we cannot determine the cause. It seems likely, given the dominance of relativistic reflection in previous low states \citep{Parker14_mrk335, Gallo15}, that there is still a strong contribution from this emission. Similarly, given the drop in flux and appearance of new absorption features in the HST spectrum, it is reasonable to expect that additional absorption will have appeared in the X-ray band as the ionisation of gas along the line of sight drops. If the source remains at these low flux levels for a longer period, we should eventually see the distant reflection drop as well, which may reveal the continuum again and allow us to determine the contribution of these different components.

The velocity we find for the warm absorber ($-5700_{-170}^{+200}$) is remarkably similar to that found by \citet{Longinotti19} ($-5700_{-400}^{+800}$), and is presumably coming from the same outflow. This velocity does not perfectly overlap with the new absorption found in the COS spectra, although it overlaps with the blue wing of the absorption troughs, but as we cannot constrain the continuum absorption in the X-ray spectra it is entirely possible that there is an X-ray counterpart to the UV absorption that does not strongly affect the emission lines. The UV absorption may also be most sensitive to the base of the outflow where the density is likely highest, while the X-ray lines are viewed through the wind further out where it is more highly ionised and faster.

% \red{Origin of line changes}
The variability of the soft emission lines discussed in section~\ref{sec_rgs} is primarily produced by a change in the ionisation or column of the warm absorber. 
Our fitting results show that changes in the emission spectra have a negligible effect on the line ratios in the O triplet in this case. The outflow velocity of the wind in \src\ (-5700~km~s$^{-1}$) is such that it aligns the 22.0~\AA\ O\textsc{vi} absorption line with the O\textsc{vii} resonance emission line, so minor variations in the absorption properties can have a large effect on the shape of the emission line. This requires a specific velocity of the wind to align the O\textsc{vi} absorption and O\textsc{vii} emission, but we note that this velocity is the same as has been observed previously for the wind in Mrk~335 in both the X-ray and UV bands \citep{Longinotti13}.
We consider an ionisation drop to be the most likely explanation for the change in wind properties, as the scale of the absorption must be large in order to cover the X-ray emitting lines, so global changes in column density are hard to achieve on short timescales (unlike changes in continuum absorption, which can occur rapidly due to clumps in the wind \citep[see e.g.][]{Risaliti07}. The ionisation, on the other hand, should vary in response to primary continuum on much shorter timescales, as ionisation propagates at the speed of light, rather than the speed of the outflow. We also note that changes in ionisation can result in a difference in the measured column density for a particular ionisation value or range while the overall column remains constant.

We note that with a single epoch spectrum, the effect of the absorption and emission could not have been disentangled, and would affect any physical properties estimated from the line ratios. This is unlikely to be an issue in type 1 AGN in general, where the continuum flux is usually strong enough for warm absorber properties to be established. However, it is worth considering for type 2 AGN, where it may be impossible to quantify the effect of warm absorption on the line ratios. This adds a degree of systematic uncertainty to estimates of the density and temperature, as well as the relative contributions of photoionised and collisionally ionised gas. As an example, naively using the lines ratios in the O triplet and the relations given in \citet{Porquet00} we would infer an order of magnitude change in the density of the emitting material, rather than a small change in the ionisation or column density of the warm absorber.

% \red{Disagreement between RGS and broadband fit results}
There is some disagreement between the best fit results between the RGS and broadband (RGS, EPIC-pn and \nustar ) spectra, in particular in the velocities of the photoionised emission components. In the RGS best-fit, the velocities are -1200~km~s$^{-1}$ and 0~km~s$^{-1}$ for the hot and cold components, respectively, while the broad-band fit prefers more intermediate values of -400 and -200~km~s$^{-1}$. In general, we consider the results from the RGS fit more reliable. The difference between the two is presumably driven by the EPIC-pn spectrum. While the pn has much higher signal and therefore dominates the fit statistic, it does not have the high resolution capabilities of the RGS, so it cannot reliably determine the velocity of narrow absorption and emission lines. We note that the uncertainty in line energies with the EPIC-pn is $\sim12$eV, due to the calibration of the gain and charge transfer inefficiency\footnote{\url{http://xmm2.esac.esa.int/docs/documents/CAL-TN-0018.pdf}}. At 1~keV, this corresponds to a velocity error of $3600$~km~s$^{-1}$, so this is to be expected.

Assuming that the velocity broadening of the photoionised emission lines measured in Section~\ref{sec_rgs} corresponds to the Keplerian velocity of the gas, we can infer the orbital radius of this material. For the cold component we estimate $R=2.4\times10^{18}$~cm (0.8~pc), and for the hot component $R=3.5\times10^{17}$~cm (0.1~pc), assuming a mass of $2.6\times10^7M_\odot$ \citep{Grier12}. This is a larger radius for the hot component than found by \citet{Longinotti08}, who constrained the density and hence distance directly by fitting the line ratios.
This is consistent with the lower velocity we measure  ($1000\pm200$~km~s$^{-1}$, rather than $2200\pm750$~km~s$^{-1}$) and the mass estimate we use is larger \citep[$2.6\times10^7M_\odot$, rather than $1.4\times10^7M_\odot$ from][]{Peterson04}. 
While the density parameter in our fits gives only a weak upper limit, we can use the bolometric luminosity ($\sim10^{44}$~erg~s$^{-1}$, from fitting the SED) along with the ionisation parameter and radius to estimate the density of the gas ($n=L/\xi R^2$). We find densities of $\sim2\times10^{6}$~cm$^{-3}$ and $\sim3\times10^{6}$~cm$^{-3}$ for the cold and hot components, respectively. This is much lower than the values estimated by \citet{Longinotti08} ($\sim10^{10-11}$~~cm$^{-3}$) for the density of the X-ray emitting clouds. We also note that the O triplet in the spectrum of \src\ from 2007 examined by \citeauthor{Longinotti08} was dominated by the intercombination line, indicating emission from high density gas \citep{Porquet00}, and the distances also much smaller (10$^{16-17}$~cm). It is therefore likely that we are seeing different material in the new spectra. One possibility is that the higher density gas lies closer to the AGN, and therefore has had time to respond to the lower source flux. In this scenario, in the current protracted low state we can only see material at larger radii, while the 2007 spectrum was more sensitive to brighter material from closer in. 

It is also possible that the velocities measured for the emission lines do not correspond to the Keplerian velocities. For example, if the AGN is viewed close to face on and the emitting region is fairly flat, most of the orbital velocity would be perpendicular to the line of sight, so we would not observe it. This would allow the emitting region to be more compact, so the absorption could also be located closer to the black hole. However, the fact that the spectrum is still dominated by photoionised emission means in 2019, at least 6 months after the descent into the low flux state, means that the line emitting region must be located at least 6 light months (0.1~pc) from the X-ray source, so we can discount this possibility. The broad iron line measured in other observations \citep{Parker14_mrk335,Gallo15,Gallo19} also argues against a fully face-on configuration, as this would produce a relatively narrow line \citep[although the inner disk does not necessarily have to be aligned with the larger AGN system, e.g.][]{Middleton16_inclinations}. Similarly, we cannot distinguish between rotational broadening and velocity gradients within the outflow itself, so the distance estimates from the velocity should be treated with caution. In principle, the most reliable estimate is from the light travel time, but this only gives a relatively weak lower limit of 0.1~pc. Further observations of Mrk~335 may be able to better constrain the distance to the photoionised material by directly measuring this delay.

An alternative way of measuring the distances is using the density of the photoionised gas. However, this is only weakly constrained, and therefore cannot be used to obtain a meaningful alternative distance measurement. The density is one of the parameters of the \textsc{pion} model, so it is estimated by spectral fitting and largely determined by the line ratios in He-like triplets. Since the O triplet is strongest, and the intrinsic line ratios are extremely uncertain because of absorption, it is no surprise that the constraints are so weak.

The clear distinction in ionisation and velocity for the two photoionised components (and in the inferred radii) means that we may be seeing separate emission from the outer X-ray BLR and X-ray NLR. The line variability is predominantly seen in the resonance line of the O\textsc{vii} triplet, which is produced by the lower ionisation component. Our spectral fitting clearly shows that this variability is produced by increased absorption, which means that the outflow must cover a large fraction ($\gtrsim0.5$) of the X-ray NLR and we must be seeing it on a very large scale. This is strong evidence that in this case the warm absorption is not produced by BLR clouds, as suggested in other sources \citep[e.g.][]{Risaliti11}. In general, the variable absorption events used to argue for occultation by BLR clouds are more commonly seen in Sy~2s, which are viewed more edge on and may be more prone to such obscuration. 
We note that the absorption in Mrk~335 is complex, so it is entirely possible that some of the warm absorption seen in higher flux states (where a more complete picture of the absorption can be obtained) is due to obscuration by BLR clouds.

Obscuration of the X-ray BLR/NLR by an outflow has not previously been observed, and is somewhat surprising, as the NLR is usually assumed to be very large compared to the size scales of outflows. However, we note that \emph{very inner} (optical) NLR in NGC~5548 was found to be only 1--3~parsecs \citep{Peterson13},  much smaller than expected \citep[although the whole NLR is widely extended and spatially resolved out to at least 900~pc, e.g.][]{Schmitt03}, and NGC~5548 is a factor of 3 more massive than Mrk~335, so the NLR in Mrk~335 could conceivably be relatively compact. We also note that there are several measurements of warm absorption size scales that put them far outside this radius \citep[see e.g.][although we note that these are generally lower velocity than the wind observed in Mrk~335]{Kaastra12, Tombesi13, Laha16}. Since the X-ray NLR is likely to be coincident with or smaller than the inner part of the optical NLR, it is therefore not impossible that an outflowing warm absorber could cover a large part of the X-ray BLR/NLR.

% \red{Geometry}
The distant reflection component is consistent between the 2018 and 2019 spectra, and dominates the spectrum above $\sim5$~keV in both models. Since we argue that the flux drop is in general attributable to intrinsic variability of the source, this means that the light-travel time to the distant reflector must be large enough that it has not yet had time to respond fully to the low flux period. This means that this material must be located at least 6 light-months (0.15~pc, $\sim2\times10^5$ gravitational radii) from the X-ray source. Results from \chandra\ spatially resolved spectroscopy of Sy~2 galaxies frequently shows that Fe~K$\alpha$ emission and associated Compton hump is produced on scales of 10s to 100s of parsecs \citep[e.g.][]{Marinucci17}. It is presumable the same large scale structure seen in Sy~2s that we are seeing with these observations, so in practical terms this represents a floor below which the X-ray flux is unlikely to drop (at least on observable timescales). Of course, it is also entirely possible that there is a contribution to the neutral reflection spectrum from more centrally located material (such as the outer accretion disk) when the central source is brighter. 

We can also estimate the covering fraction of the reprocessor: if we assume that the central X-ray source emits isotropically, as does the reprocessing medium, then the covering fraction will be equal to the fraction of the flux that is reprocessed into the neutral reflection spectrum. Assuming that the average long term 2--10~keV flux of \src\ is $\sim1\times10^{-11}$~erg~s~cm$^{-2}$ \citep[see long-term lightcurve in][]{Parker14_mrk335}, combining this with the 2--10~keV flux of $\sim5\times10^{-13}$~erg~s~cm$^{-2}$ for the reflection component in our models gives an approximate covering fraction of $\sim5\%$. This is obviously a very uncertain value, but it is comparable with the small ($\lesssim10\%$) covering fraction for this material inferred in Sy~2 NGC~4945 based on variability arguments \citep[e.g.][]{Yaqoob12, Puccetti14}.

%%%%%%%%%%%%%%%%%%%%%%%%%%%%%%%%%%%%%%%%%%%%%%%%%%

\section{Conclusions}
We have presented new \xmm , \nustar, \swift\ and HST observations of the NLS1 Mrk~335 at its lowest flux level yet, which has lasted for at least 6 months. This is a very rich dataset, which we will explore further in future work. Our initial findings, based primarily on X-ray spectroscopy, are:

\begin{itemize}
    \item The high resolution RGS spectra in 2018 and 2019 are well described with a power-law continuum, two photoionised emission components, which we attribute to the outer X-ray BLR and X-ray NLR, and a single layer of ionised absorption. The dramatic change in O\textsc{vii} line ratios between the two spectra can be attributed to a drop in ionisation of the absorber, which strongly absorbs the resonance line.
    \item The photoionised emission is consistent with either remaining constant or dropping slightly in ionisation. Lower ionisation values are found for both components in 2019 in all fits, but the difference is small and the uncertainties large.
    \item The broad-band \xmm/\nustar\ X-ray spectrum is dominated at low energies by the photoionised emission, and at high energies by distant reflection. The dominance of these components is likely responsible for the lack of X-ray variability observed by \xmm\ and \swift , while the UV continues to vary as normal.
    \item It is also clear from the broad-band fitting that absorption cannot be the primary driver of the flux drop, as it does not explain the relatively uniform drop in flux across the energy band or the simultaneous drop in UV continuum flux.
    \item The lack of response from the photoionised lines and distant reflection implies that these components originate more than 6 light-months from the source, and thus have not yet had time to respond to the 2018 flux drop. This is confirmed by fitting the velocity broadening of the lines in the RGS spectrum, which gives radii of $\sim$0.1 and 0.8~pc for the hot and cold photoionised gas. 
    \item It follows from the varying obscuration of these lines that the outflowing warm absorber must extend at least this far from the X-ray source as well, otherwise it could not affect the scattered emission. This means that the absorption must be associated with a large-scale outflow, rather than BLR clouds, and the observed outflow velocity of the absorber does not correspond to the Keplerian velocity.
\end{itemize}

Mrk~335 remains a fascinating source for study, exhibiting almost every phenomenon observed in AGN X-ray spectroscopy, often at the same time. The spectra presented here demonstrate that triggered observations of AGN low states may be the best way of understanding the nuclear environment of type 1 AGN in the X-ray band.

%%%%%%%%%%%%%%%%%%%%%%%%%%%%%%%%%%%%%%%%%%%%%%%%%%

\section*{Acknowledgements}
MLP, GAM and CP are supported by European Space Agency (ESA) Research Fellowships. ACF acknowledges support from ERC Advanced Grant 340442. ALL is supported by CONACyT grant CB-2016-01-286316. JJ Acknowledges support from the Cambridge Trust and the Chinese Scholarship Council joint scholarship (201604100032). DG acknowledges support by HST grant HST-GO-15439.002-A. Based on observations obtained with XMM-Newton, an ESA science mission with instruments and contributions directly funded by ESA Member States and NASA. This work made use of data from the \nustar\ mission, a project led by the California Institute of Technology, managed by the Jet Propulsion Laboratory, and funded by the National Aeronautics and Space Administration. This research has made use of the \nustar\ Data Analysis Software (NuSTARDAS) jointly developed by the ASI Science Data Center (ASDC, Italy) and the California Institute of Technology (USA). We acknowledge support from the Faculty of the European
Space Astronomy Centre (ESAC). We thank the anonymous referee for their constructive feedback, which has significantly improved this work.

%%%%%%%%%%%%%%%%%%%%%%%%%%%%%%%%%%%%%%%%%%%%%%%%%%



\bibliographystyle{mnras}
\bibliography{bibliography} % if your bibtex file is called example.bib

%%%%%%%%%%%%%%%%%%%%%%%%%%%%%%%%%%%%%%%%%%%%%%%%%%

%%%%%%%%%%%%%%%%% APPENDICES %%%%%%%%%%%%%%%%%%%%%
\appendix
\section{PION\_XS Table Model}
\label{sec_pionxs}

In its analytical form \textsc{pion} is extremely computationally expensive, so even relatively simple fits take a very long time. The conventional approach within X-ray astronomy for complex models has been to pre-calculate grids of model spectra, then interpolate between grid points when fitting, as is the case with \textsc{reflionx} \citep{Ross05}, \textsc{xillver} \citep{Garcia13}, \textsc{kerrbb} \citep{Li05} or \textsc{xstar} \citep{Kallman01}.

Whether this is appropriate or useful depends on the circumstances. \citet{Kaastra16} discuss the use of pre-calculated table models as an alternative to optimised response matrices and spectral binning to improve the speed of spectral fitting. They conclude that it is too computationally expensive to calculate grids covering the full parameter space relevant for astronomy, and it is preferable to provide an analytic model and optimise the fitting procedure as much as possible. It is also the case that a table model will be less accurate than an analytical model, as it relies on interpolation to estimate the spectrum between grid points \citep[one case of the model interpolation giving unreliable results is discussed in the appendix of][]{Parker19_continuum}. 

In general, as discussed by \citeauthor{Kaastra16}, model grids shift the computing burden from the fitting procedure to the grid creation. However, this is not an equal trade-off, since a grid model must only be evaluated once for each grid point, whereas an analytical model must be evaluated at every point while fitting, for every fit run by every user. 

\begin{table}
    \centering
    \caption{Parameter ranges for the preliminary \textsc{pion\_xs} grid used to model the RGS emission lines.}
    \label{tab:gridpars}
    \begin{tabular}{lcccr}
    \hline
    \hline
    Parameter       &   Min &   Max    &    $N$     & Unit\\
    \hline
    $\log(\xi)$     &   -2  &   3       &   11      & erg~s~cm$^{-2}$\\
    Density         &   $10^8$   &   $10^{12}$   &   9   & cm$^{-3*}$\\
    \hline
    \end{tabular}
    
    $^*$Note that the unit for number density in \textsc{spex} is 10$^{20}$~m$^{-3}$, which we convert to cm$^{-3}$ for consistency with the rest of the paper.
\end{table}

In this case, we are interested in a single source with a relatively constant spectral shape, where the physical conditions of the emitting gas are unlikely to change drastically. We also require a model that can be used with the wider range of models available within \textsc{xspec}, and directly converting the \textsc{pion} code to an \textsc{xspec} model would be extremely difficult.
We therefore calculate \textsc{pion\_xs}, an \textsc{xspec} table version of \textsc{pion}. For simplicity, we only consider the emission spectrum. Using the ESAC science grid, we evaluate \textsc{pion} over a grid of ionisation and electron density values, assuming an input $\Gamma=2$ powerlaw spectrum and a fixed column density of 10$^{20}$~cm$^{-1}$, on the ESAC science grid. We save the output spectra, and convert these to an \textsc{xspec} table model. The variable parameters used for the grid are given in Table~\ref{tab:gridpars}. The RMS velocity is fixed at 100~km~s$^{-1}$, and the atomic abundances are fixed at solar, using the values from \citet{Lodders09}. 

This conversion dramatically increases the speed with which fits can be performed, and means that \textsc{pion} can be used in conjunction with the more varied models available in \textsc{xspec}, but limits the precision of the model and the physical cases that it is appropriate for. 

\begin{figure}
    \centering
    \includegraphics[width=\linewidth]{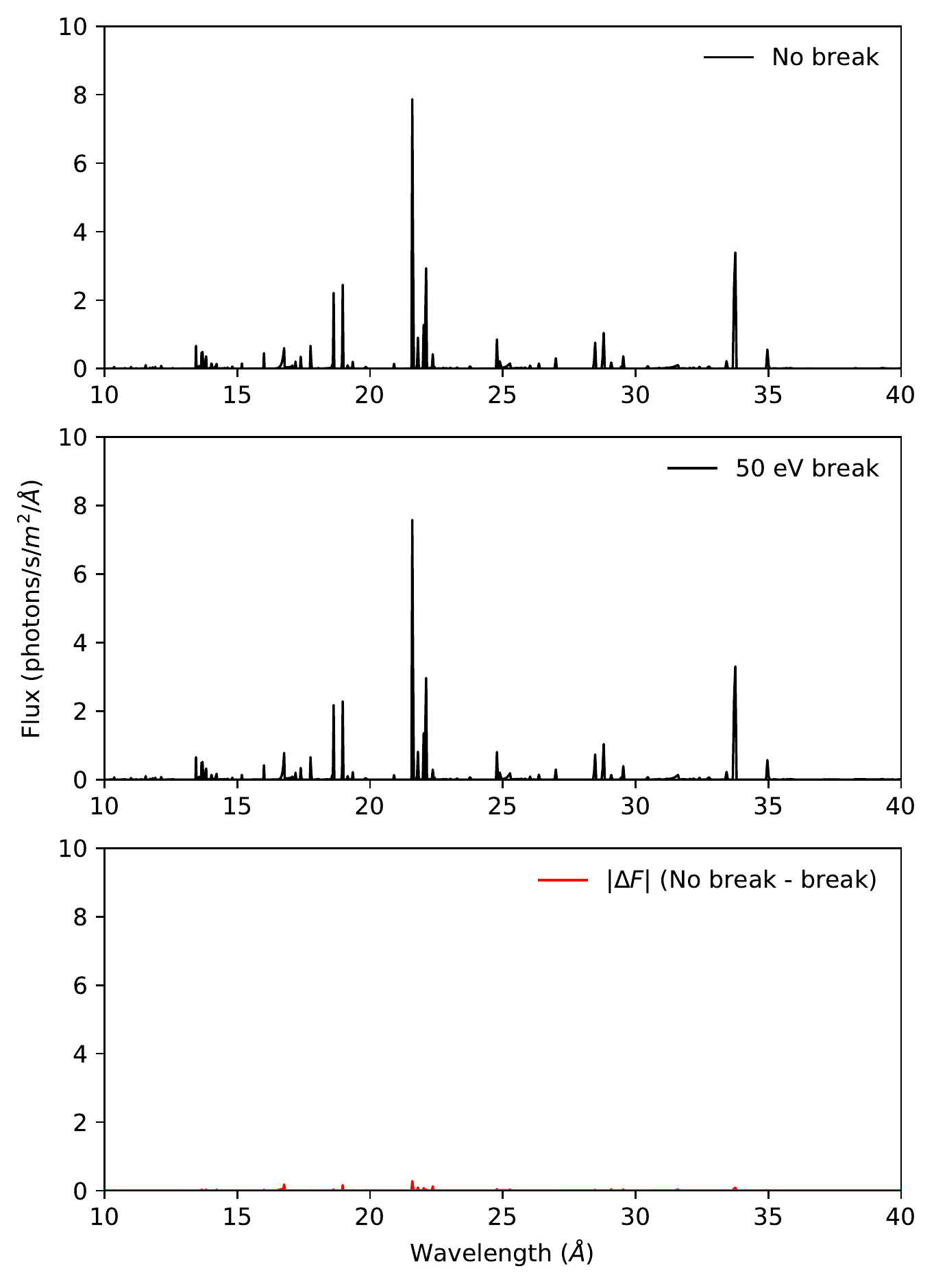}
    \caption{Top: \textsc{pion} emission line spectrum, assuming an unbroken power-law input spectrum. Middle: as above, but with a sharp break at 50~eV. Bottom: the absolute difference between the two spectra.}
    \label{fig:break_comparison}
\end{figure}

One important consideration is that our treatment of the continuum model is simplistic. We assume an unbroken $\Gamma=2$ power-law, extending indefinitely into the UV band. This means that the power-law contributes a lot of UV photons in addition to the X-rays, which can potentially significantly impact the ionisation balance of the gas. We explore the likely impact of this by simulating a single spectrum with a sharp break at 50~eV (breaking to $\Gamma=-8$ below this energy), so that there is no contribution from UV photons. For the other parameters, we assume an ionisation of $\log(\xi)=1$ erg~s~cm$^{-2}$ and a density of $10^{-8}$~cm$^{-3}$. The two spectra, and the difference between them, are shown in Fig.~\ref{fig:break_comparison}. They are almost indistinguishable, so we conclude that the assumption of a pure power-law continuum is reasonable for the case of pure line emission.

The \textsc{pion\_xs} table can be downloaded from \href{http://bit.ly/pion_xs}{bit.ly/pion\_xs}. 

%%%%%%%%%%%%%%%%%%%%%%%%%%%%%%%%%%%%%%%%%%%%%%%%%%

%%%%%%%%%%%%%%%%%%%%%%%%%%%%%%%%%%%%%%%%%%%%%%%%%%

% Don't change these lines
\bsp	% typesetting comment
\label{lastpage}
\end{document}